\PassOptionsToPackage{hidelinks,colorlinks,linkcolor={blue!50!black},urlcolor={blue!50!black},citecolor={blue!50!black}}{hyperref}
\pdfoutput=1
\documentclass[12pt,a4paper]{article}
\usepackage{url}
\usepackage{graphicx}
\usepackage[usenames]{xcolor}
\usepackage{hyperref}
\usepackage{t1enc}
\usepackage[utf8]{inputenc}
\usepackage[english]{babel}
\usepackage{amsmath,amssymb,amsthm}
\newtheorem{thm}{Theorem}

\newtheorem{cor}{Corollary}
\newtheorem{exa}{Example}
\theoremstyle{definition}
\newtheorem{remark}{Remark}
\theoremstyle{definition}

\usepackage{enumitem}
\usepackage{booktabs}
\usepackage{xspace}
\usepackage{multirow}
\usepackage{array}
\usepackage{natbib}
\usepackage{bm}
\usepackage{setspace}
\newcommand{\appendixref}[1]{Appendix~\ref{#1}}
\newcommand{\techref}[1]{Section~\ref{#1}}
\usepackage[normalem]{ulem}
\usepackage{calc}
\newcommand{\stkout}[1]{\ifmmode\text{\sout{\ensuremath{#1 }}}\else\sout{#1}\fi}
\newcommand{\correction}[2]{#2} 

\newcommand{\hta}{\ensuremath{\text{5-HT}_{\text{2A}}}\xspace}
\newcommand{\htt}{SERT\xspace}
\newcommand{\bpnd}{\ensuremath{\text{BP}_{ND}}\xspace}
\newcommand{\bpp}{\ensuremath{{\text{BP}_{p}}\xspace}}

\newcommand{\var}{\ensuremath{\mathbb{V}\text{ar}}}
\newcommand{\cov}{\ensuremath{\mathbb{C}\text{ov}}}

\newcommand{\E}{\ensuremath{\mathbb{E}}}

\newcommand{\influ}{\ensuremath{I\!F}}

\newcommand{\twostage}{\texttt{2SSEM}\xspace}
\newcommand{\predphi}{\varphi^*} 
\usepackage{bm}
\usepackage{setspace}
\usepackage{fancyvrb}
\usepackage{verbatim}
\usepackage{listings}
\usepackage{lmodern}
\usepackage[paperwidth=210mm,
paperheight=297mm,
marginparsep=15pt,
marginparwidth=7em,
]{geometry}

\begin{document}

{\huge \bf
  \begin{center}
    A two-stage estimation procedure for 
    non-linear structural equation models
  \end{center}} 
\vspace{0.2 cm}

\begin{center}
  {\large Klaus Kähler Holst\({}^{123*}\), Esben Budtz-Jørgensen\({}^{3*}\)}\\
\end{center}

{
  \footnotesize
  \emph{\(^{1}\) Maersk Analytics,
    Esplanaden 50, 
    DK-1098 Copenhagen K, Denmark.}\\
  \emph{\(^{2}\) Neurobiology Research Unit, Rigshospitalet,
    Copenhagen University Hospital.
    Juliane Maries Vej 28, building 6931, 3rd floor, DK-2100 Copenhagen, Denmark.} \\
  \emph{\(^{3}\)Department of Biostatistics,  University of Copenhagen.
    Øster Farimagsgade 5, entr. B, 
    P.O.Box 2099, DK-1014 Copenhagen K, Denmark.}\\
  \emph{\({}^{*}\)} Authors contributed equally
}

\vspace{0.5 cm}

\begin{center}
  SUMMARY
\end{center}
Applications of structural equation models (SEMs) are often restricted
to linear associations between variables. Maximum likelihood (ML)
estimation in non-linear models may be complex and require numerical
integration.  Furthermore, ML inference is sensitive to distributional
assumptions.  In this paper, we introduce a simple two-stage estimation
technique for estimation of non-linear associations between latent
variables.  Here both steps are based on fitting \emph{linear} SEMs:
first a linear model is fitted to data on the latent predictor and
terms describing the non-linear effect are predicted by their
conditional means.  In the second step, the predictions are included
in a linear model for the latent outcome variable. We show that this
procedure is consistent and identifies its asymptotic distribution.
We also illustrate how this framework easily allows the association
between latent variables to be modelled using restricted cubic splines
and we develop a modified estimator which is robust to non-normality
of the latent predictor. In a simulation study, we compare the proposed
method to MLE and alternative two-stage estimation techniques.

\section{Introduction}

Over the last decades linear structural equation models have been useful in
many fields of research. These models typically consists
of two parts, a measurement part where observed outcomes are
assumed to be reflections of underlying latent variables and a
structural part relating the latent variables to each other. 
Important extensions of this framework have used more flexible measurement
models to allow  inclusion of binary,
ordinal and censored outcomes (\cite{muthen1984general},
\cite{skrondal2004generalized}). In this paper
we focus on the structural part 
and consider models which allow for non-linear relations between
the latent variables. Until now most research in this \correction{framework}{topic} has
been on the interaction model assuming a linear effect of the product
of two latent variables (i.e., \(\eta=\beta_1 \xi_1+ \beta_2
\xi_2+\beta_3 \xi_1 \xi_2 + \zeta\)) or more general polynomial
models including terms of higher order (e.g., \(\eta=\beta_1 \xi_1+ \beta_2
\xi_1^2 + \zeta\)), but a more general framework is of obvious interest. 

Maximum likelihood inference in linear SEMs are facilitated by the
fact that a closed form expression for the likelihood function is
obtained when integrating out the latent variables.  Non-linear models
do not have this property and numerical methods are needed. Today, the
so-called LMS algorithm \citep{klein00:_maxim_likel_estim_lms} is
perhaps the most widely used method. It approximates the likelihood
function using a mixture of multivariate normal distributions and then
this function is maximized with the EM-algorithm. For simple
non-linear models (product-interaction model and second degree
polynomial), this algorithm has been implemented in the widely used
software package \emph{Mplus} \citep{muthen2012mplus}.
\correction{An alternative EM-algorithm not
based on the mixture approximation was proposed by Lee and Zhu (2002), and Rizopoulos and
Moustaki (2008) suggested a hybrid EM algorithm using Adaptive Gaussian Quadrature (AGQ).
AGQ was also used by Wall (2009).}{An alternative
EM-algorithm was proposed by
\cite{lee2002maximum}, while \cite{wall2009maximum} used  Adaptive Gaussian
Quadrature (AGQ)
and \cite{rizopoulos_glvm_2008} suggested a
hybrid EM-algorithm based on AGQ.} 
Bayesian techniques have also been considered for non-linear models
(\cite{arminger1998bayesian}, \cite{lee2000statistical}) and these
have been extended to more flexible semi-parametric
models (\cite{yang2010bayesian}, \cite{song2010semiparametric}, \cite{kelava2014general}).

Due to the complexity of the ML-procedure for estimation of non-linear
SEMs, a number of simpler so-called limited information methods have
been developed. \cite{kenny1984estimating}
developed the first estimator for the product-interaction model by
fitting a modified linear model including an additional latent
variable with indicators given by products of indicators of the two
interacting latent variables. Since then this technique has been
refined by several researchers (see \cite{marsh2004structural} for an
overview), but it remains rather ad-hoc and cannot be used for more
general non-linear models. For the polynomial model, 
\cite{wall2000estimation} proposed a two-step method (2SMM), where the latent
variables were first predicted using the so-called Bartlett score,
while the second step estimated the parameters of the non-linear
relations using a method-of-moments procedure allowing for uncertainty
in the predicted variables. Despite some nice statistical properties,
this method has not been used much in practice. \cite{mooijaart2010alternative}
developed a method-of-moments procedure by including third degree
moments for estimation of non-linear effects in the polynomial model. A computationally very simple
two-stage least squares (2SLS) method was developed by Bollen
(\cite{bollen1995structural}, \cite{bollen1998interactions}). Here
instrumental variables must be identified also for non-linear
terms. It is not clear whether this method can be used in general
models, but the method is non-iterative and easy to implement in
standard statistical software. However, simulation studies have
indicated
that this method has a rather low efficiency compared to ML-estimation
\citep{schermelleh1998estimating}. 

In this paper we present a new two-stage method for estimation in
non-linear structural equation models. As in \correction{the Wall-Amemiya
procedure}{2SMM}, we first have a prediction step, but instead of the
Bartlett score we use the Empirical Bayes method and instead of
predicting the latent variables we predict the latent non-linear
effect terms. Therefore, in the second step, it is sufficient to fit a
linear structural equation model with the predicted variables included
as covariates.  We show that the method yields consistent estimation
and derive expressions for asymptotic standard errors. We illustrate
how splines can be included and by using mixture models in the first
step, we extent the method so that it becomes robust to non-normality
of latent predictor variables. In simulation studies the method is
compared to ML, \correction{and Bollen's}{2SMM, 2SLS and an
  alternative two-stage estimator of semi-parametric associations 
between latent
  variables \citep{kelava2017}}. Finally, we illustrate
the usefulness of the method analysing data from neuroscience on the
regional binding potential of different serotonergic markers in the
human brain.

\section{A non-linear structural equation model}\label{sec:nonlin}
We consider a model where a latent response variable 
\(\eta_i=(\eta_{i1},...,\eta_{ip})^t\) of subject \(i\)
(\(i=1,\ldots,n\)) is assumed to be non-linearly
related to a latent predictor \(\xi_i=(\xi_{i1},...,\xi_{iq})^t\) 
after adjustment for covariates $Z_i=(Z_{i1},...,Z_{ir})$
\begin{eqnarray}\label{stmodel}
  \begin{array}{lcl} 
    \eta_i & = & \alpha + B \varphi(\xi_i) + \Gamma Z_i + \zeta_i,
  \end{array} 
\end{eqnarray}
such that
\(\varphi(\xi_i)=(\varphi_1(\xi_{i}),...,\varphi_l(\xi_{i}))^t\) has finite
variance. The main parameter \(B (p \times l)\) describes
the non-linear relation between \(\xi_i\) and \(\eta_i\).
 Note, that \(\varphi\) may also depend on some of the
covariates thereby allowing the introduction of interaction terms, but we here omit this to simplify notation.

The latent
predictors
(\(\xi_i\))
are related to each other and the covariates in a  linear structural
equation 
\begin{eqnarray}\label{stmodel1}
  \begin{array}{lcl} 
    \xi_i & = & \tilde{\alpha} + \tilde{B} \xi_i+\tilde{\Gamma} Z_i + \tilde{\zeta}_i,
  \end{array} 
\end{eqnarray}
where diagonal elements in  \(\tilde{B}\) are zero and
the residual terms \(\zeta_i\) and  \(\tilde{\zeta}_i\) are
assumed to be independent with mean 0 and 
covariance matrices of \(\Psi\) and \(\tilde{\Psi}\), respectively. 

The observed variables \(X_i=(X_{i1},...,X_{ih})^t\) and
\(Y_i=(Y_{i1},...,Y_{im})^t\) are linked to the latent variable 
in the two measurement models
\begin{align}
  \label{memodely}
  Y_i  =  \nu + \Lambda \eta_i  + K Z_i+\epsilon_i \\
  \label{memodelx}
  X_i =  \tilde{\nu}+\tilde{\Lambda} \xi_i + \tilde{K} Z_i+ \tilde{\epsilon}_i,  \end{align}
where the error terms \(\epsilon_i\) and  \(\tilde{\epsilon}_i\) are
assumed to be independent  with mean 0 and 
covariance matrices of \(\Omega\) and \(\tilde{\Omega}\), respectively.
The parameters are collected into \(\theta=(\theta_1,
\theta_2)\), where \(\theta_1=(\tilde{\alpha}, \tilde{B},
\tilde{\gamma},
\tilde{\nu}, \tilde{\Lambda}, \tilde{K},
\tilde{\Omega},\tilde{\Psi})\) are the parameters of the linear SEM 
describing the conditional distribution of \(X_i\) given \(Z_i\). The
rest of the parameters are collected into \(\theta_2\).

For identification of the model, we need to impose some parameter
constraints \citep{bollen1989structural}. Generally, measurement models
can be made identifiable by selecting a reference indicator for each
latent variable. For this variable we fix the regression coefficient
of the latent variable
(element of \(\Lambda\) or  \(\widetilde{\Lambda}\)) to 1 and the intercept
(element of \(\nu\) or \(\widetilde{\nu}\)) to 0. Alternatively, the
variance of latent variables can be fixed to 1, and their intercepts
(element of \(\alpha\) or \(\widetilde{\alpha}\)) set to 0. In the estimation
procedure described below, it turns out to be crucial to use the
reference indicator restriction in the measurement model for $Y_i$ . Also, we model the covariance \(\Psi\) of the latent outcomes (\(\eta_i\))
using an unrestricted covariance matrix.

\begin{figure}[htbp!]
  \centering
  \includegraphics[width=0.5\textwidth]{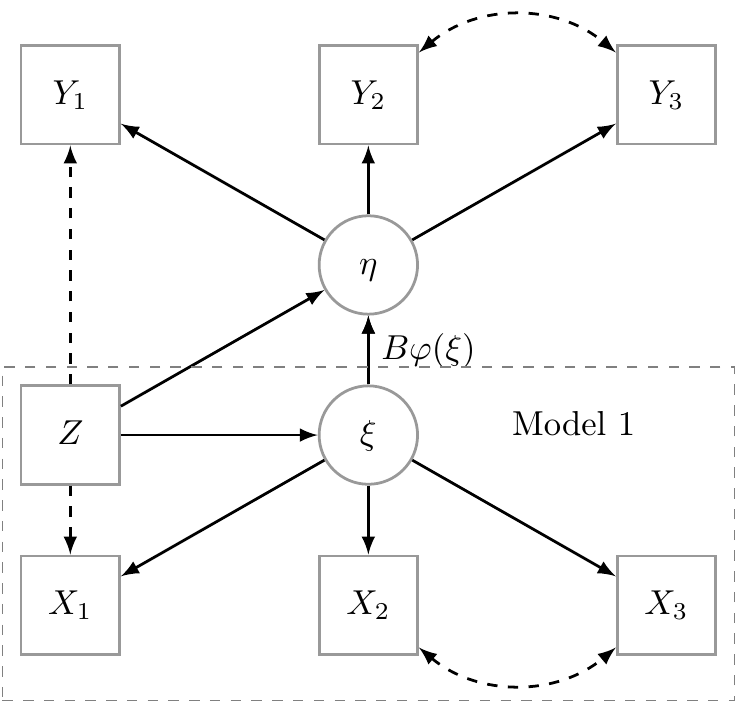}
  \caption{Path diagram showing an example of the non-linear structural equation
    models considered. The
    two-stage estimator is constructed by separately estimating
    parameters of Model 1, and (non-linear) associations between the two
     models are then estimated in a subsequent step
    based on predictions from the Model 1 analysis.
    \label{fig:path1}}
\end{figure}

The likelihood function is \(L(Y,X,Z,\theta)= \prod_{i=1}^{n}\int \, p_{\theta}(Y_i,X_i \vert
    \eta_i, \xi_i,Z_i) p_{\theta_2}(\eta_i \vert
    \xi_i, Z_i) p_{\theta_1}(\xi_i \vert Z_i)  d \xi_i d \eta_i\).
 Assuming joint normality of
\((\zeta_i, \tilde{\zeta}_i, \epsilon_i,\tilde{\epsilon}_i)'\), a closed
form expression for \(L\), is available only
if \(\varphi\) is linear. In the general case numerical integration
techniques are necessary for ML-estimation, which in practice for even
moderately sized problems (number of latent variables) is
computationally intractable.  \correction{First, we note that from the
  structural equation (\ref{stmodel}) we get}{Instead, we exploit  that the structural
equation (\ref{stmodel}) is linear in the parameters to get}
\begin{eqnarray}\label{idea}
  \begin{array}{lcl}
    \E(\eta_i \vert X_i, Z_i)  =  \alpha + B  \E_{\theta_1} [\varphi(\xi_i)\vert 
    X_i,Z_i] + \Gamma Z_i,
  \end{array} 
\end{eqnarray}
noting that the conditional expectation on the right-hand side depends only on the
distribution parametrized by \(\theta_1\).
Equation (\ref{idea}) suggests that parameters can be estimated in two steps. First, 
the linear SEM given by equations (\ref{stmodel1}) and 
(\ref{memodelx}) 
is fitted to \((X_i,Z_i, i=1,\ldots,n)\)  and the latent covariate \(\varphi(\xi_i)\)
is predicted by the conditional mean \(E_{\widehat{\theta}_1}[ \varphi(\xi_i)\vert X_i,Z_i]\). 
Step 2 then estimates \(\theta_2\) in a linear SEM, where the
measurement model is given by equation (\ref{memodely})
and the structural model is equation (\ref{stmodel}) with the latent
predictor replaced by \(\E_{\widehat{\theta}_1}[\varphi(\xi_i) \vert X_i,Z_i]\). Thus, the key idea is to
replace  \(\varphi(\xi_i)\) 
by (an estimate of) the conditional mean of \(\varphi(\xi_i)\) given \(X\)  and
\(Z\). Previous
methods have used \(\varphi\) of the conditional mean
\(E_{\widehat{\theta}_1}[\xi_i \vert
X_i,Z_i]\) (\cite{joreskog2000latent}, \cite{schumacker2002latent}).

In the following we will use the notation
\(\predphi(\xi_i) = \E[\varphi(\xi_i)\mid X_i,Z_i]\) and
\(\predphi_n(\xi_i) = \E_{\widehat{\theta}_1}[\varphi(\xi_i)\mid X_i,Z_i]\) to
distinguish between the conditional expectation and the
plug-in estimator, where the expectation is taken with respect to the
distribution indexed by the estimated parameter values from the stage
one model.
To summarize, we define the two-stage estimator in the following way

\begin{remark}[\correction{Two-stage estimator}{Two-Stage Structural
    Equation Model estimator (\twostage)}]\label{def:twostage}~ 
  \begin{enumerate}
\item  The linear SEM given by equations (\ref{stmodel1}) and is
  (\ref{memodelx}) 
  is fitted to \((X_i,Z_i, i=1,\ldots,n)\) to achieve an estimate of
  the parameter  \(\theta_1\).
\item   The parameter \(\theta_2\) is estimated via a linear SEM with
  measurement model given by equation (\ref{memodely})
  and structural model given by equation (\ref{stmodel}), where the latent
  predictor, \(\varphi(\xi_i)\), is replaced by \(\predphi_n(\xi_i)\).
\end{enumerate}
\end{remark}


We can now formulate the consistency properties of the proposed
estimator (the asymptotic distribution of \twostage is
derived in \techref{sec:asym}).

\begin{thm}\label{thm:consistency}
  Under a correctly specified non-linear SEM
  (\ref{stmodel})-(\ref{memodelx}) including correctly specified
  distribution of the residual terms, \twostage  will yield consistent estimates of all
  parameters (\(\theta\)) except for the residual covariance,
\(\Psi\), of the latent variables in step 2.

  \begin{proof}
    Since the exposure part of the model is correctly specified,
    \(\theta_1\) is estimated using ML-estimation and therefore
    \(\widehat{\theta}_1\) is consistent under mild regularity
    conditions \citep{anderson_normalfactor_annals1988}. In step two, the model is misspecified as we fit a linear model to
    a non-linear association.  We prove the theorem by showing that
    even though the model is misspecified it includes the true mean
    and covariance of the data. When fitting a linear SEM with
    correctly specified mean and variance, the estimator
    \(\widehat{\theta}_2\) will converge to the parameter value which
    induces the true mean and variance
    \citep{arminger1989pseudo}. Finally, we characterize this parameter
    value and see that it is identical to the truth (\(\theta_2\)) except
    for elements describing the residual covariance of the latent
    variables in step 2 (\(\Psi\)). We make these arguments assuming that
    the predicted values \(\predphi\) were available.  The
    result then follows from the Continuous Mapping Theorem and by
    noting that \(\predphi_n\longrightarrow\predphi \ a.s.\) as
    \(n\to\infty\). The proof is illustrated in the situation where
    \(K=0\), as this simplifies matrix expressions without affecting
    theoretical insights
    (see the appendix for a more general formulation). 

The next step is to show that the model fitted in 
step-two induces the correct
mean and variance structure for  the data
    \(V_i = (Z_i^t,\predphi(\xi_i)^t, Y_i^t)^t\). To save space,
we illustrate this only for the
    variance (full proof is given in appendix).  
    In step two, we fit a linear model with correct measurement part
    (\(Y_i = {\nu_2} + \Lambda_2 \eta_i +\epsilon_i\)) and structural part
    \( \eta_i = \alpha_2 + B_2\predphi(\xi_i) + \Gamma_2 Z_i +
    \zeta_i\). Here the subscript 2 is used to
    distinguish the parameters of step-two model from the true
    parameter value (no subscript).  Using standard
    results from linear models \citep{bollen1989structural}, and the
    notation \(\Sigma_{z,\varphi} = \Sigma_{\varphi,z}^t\),
    the modelled variance is
\begin{eqnarray}\label{varm}
 \begin{array}{ccl}
   \Sigma_{V_i}
   &
     =
   &
     \left( \begin{array}{ccc}              
              \Sigma_Z
              &
                \Sigma_{z,{\varphi}}
              &
                [
                \Sigma_{z,{\varphi}}B_2^t
                +
                
                \Sigma_z \Gamma_2^t] \Lambda_2^t\\
              . &
                  \Sigma_{\varphi}
              &
                [
                \Sigma_{\varphi} B_2^t
                + 
                \Sigma_{{\varphi},z} \Gamma_2^t
                ] \Lambda_2^t\\
              . &
                  . & \Lambda_2
                      \Sigma_{\eta}
                      \Lambda_2^t
                      +
                      \Omega_2  
            \end{array}\right)
      \end{array}
  \end{eqnarray}
\noindent where 
   $\Sigma_{\eta}=B_2 \Sigma_{\varphi} B_2^t + B_2
              \Sigma_{{\varphi},z} \Gamma_2^t +
            \Gamma_2\Sigma_{z,{\varphi}} B_2^t + \Gamma_2
            \Sigma_Z \Gamma_2^t + \Psi_2$
  is the modelled variance of \(\eta\) and \(
  \Sigma_Z,  \Sigma_{z,{\varphi}},\Sigma_{\varphi}\) are completely unstructured 
  parameters modelling the variance of the step-two covariates
  (\(Z_i,\predphi(\xi_i)\)). 
 To calculate
  the true variance, we write the structural
  model as 
  \begin{eqnarray}\label{stmodela}
    \begin{array}{l} 
      \eta_i  =  \alpha + B \varphi(\xi_i) + \Gamma Z_i + \zeta_i
 =  \alpha + B\predphi(\xi_i)   + B [\varphi(\xi_i)-
                  \predphi(\xi_i)] + \Gamma Z_i + \zeta_i,
    \end{array} 
  \end{eqnarray}

\noindent and from this we derive the variance of the latent variable
\(\eta_i\)
    \begin{eqnarray}\label{varetat}
      \begin{array}{lcl} 
        \var(\eta_i) &= & B \var [\predphi(\xi_i)] B^t + B 
                          \cov[\predphi(\xi_i), Z_i] \Gamma^t +  \Gamma \cov[Z_i,
                          \predphi(\xi_i)] B^t\\
                     & & + B \var[\varphi(\xi_i)- \predphi(\xi_i)]B^t + \Gamma \var(Z) \Gamma^t + 
                         \Psi,
      \end{array} 
    \end{eqnarray}

\noindent because
    \(\cov[\predphi(\xi_i), \varphi(\xi_i)-
    \predphi(\xi_i)]=0\) and  \(\cov[Z_i, \varphi(\xi_i)-
    \predphi(\xi_i)]=0\), which follows from  
 the law of iterated expectations.
We can now determine the covariances 
\(\cov[\predphi(\xi_i),Y_i]=\cov[\predphi(\xi_i),\Lambda \eta_i] = 
\cov[\predphi(\xi_i), B
    \predphi(\xi_i)+ \Gamma Z_i] \Lambda^t\) as \(\cov[\predphi(\xi_i), \varphi(\xi_i)-
    \predphi(\xi_i)]=0\)
    and
\(\cov(Z_i,Y_i)=\cov(Z_i\Lambda \eta_i)= \cov[Z_i,B
    \predphi(\xi_i)+ \Gamma Z_i] \Lambda^t\) as \(\cov[Z_i, \varphi(\xi_i)-
    \predphi(\xi_i)]=0\). Finally, the
    variance is 
\begin{eqnarray}\label{vart}
\hspace{-14pt}
      \begin{array}{l}
        \var(V_i) =
                                                 
                                                   \left( \begin{array}{ccc}
                                                           \var(Z_i) &
                                                           \cov[Z_i,\predphi(\xi_i)] &
                                                           [\cov\{Z_i,\predphi(\xi_i)\}
                                                         B^t+
                                                           \var(Z_i)
                                                           \Gamma^t] \Lambda^t\\
                                                            . &   \var[\predphi(\xi_i)] &  [\var
                                                            \{\predphi(\xi_i)\} B^t + 
                                                            \cov\{\predphi(\xi_i),Z_i\} \Gamma^t]\Lambda^t\\
                                                             . &  . & 
                                                            \Lambda
                                                            \var(\eta_i)
                                                            \Lambda^t
                                                            + \Omega 
                                                          \end{array}\right).
      \end{array}
    \end{eqnarray}

\noindent It can now be seen that the modelled variance
    is equal to the true variance [(\ref{varm}) =
    (\ref{vart})] and the modelled mean is equal to the true mean
    (equations not shown)
if \(B_2=B,
    \Lambda_2=\Lambda, \Gamma_2=\Gamma, \Omega_2=\Omega, \alpha_2= \alpha,
    \nu_2=\nu, \mu_{\varphi}= \E[\varphi(\xi_i)], \mu_Z=\E(Z_i),
    \Sigma_{z,{\varphi}}=\cov[Z_i,\predphi(\xi_i)],
    \Sigma_{\varphi}= \var[\predphi(\xi_i)], 
    \Sigma_Z=\var(Z_i)\) 
    and  \(\Psi_2=\Psi+B \var[\varphi(\xi_i)-
    \predphi(\xi_i)] B^t\). Note, that although the expression for
    the modelled variance of the latent variable (\(\Sigma_{\eta}\))
    is missing the term 
\(B \var[\varphi(\xi_i)- \predphi(\xi_i)] B^t\), 
the model  can achieve the correct
    variance for \(Y_i\) 
    by adding the missing term to the residual variance \(\Psi_2\). 
    Since the model
    includes the true mean and variance, \(\widehat{\theta}_2\) will
    converge to the specific parameter value 
inducing this mean and variance. Therefore, all parameters are
    consistently estimated 
    except for the variance of the latent variable,
    which will be overestimated.     
  \end{proof}
\end{thm}

An important advantage of the proposed method is that closed form expressions for
\(\predphi(\xi_i)\) are available for large classes of functions
including polynomials and splines basis functions (see
\techref{sec:prediction}) making the implementation of the estimator
straightforward and computationally fast.
Also, note that the formulation of Theorem \ref{thm:consistency}
is overly restrictive, as it
states that consistency relies not only on a correctly specified model
structure, but also on correctly specified distributions of
residuals. However, linear SEMs only require a correctly specified
mean and variance (conditional on covariates) to yield consistent
estimation \citep{arminger1989pseudo} and therefore \correction{the two-stage
estimator}{\twostage} will be robust to distributional misspecifications. Thus,
non-normality of the residuals of the measurement models in step two
will not affect consistency.  However, for \(\widehat{\theta}_2\) to be
consistent, we note that the following conditions must hold:
(a) \(\E[\varphi(\xi_i)-\varphi^*(\xi_i)]=0\),
(b) \(\cov[\varphi^*(\xi_i),\varphi(\xi_i)-\varphi^*(\xi_i)]=0\),
(c) \(\cov[Z_i,\varphi(\xi_i)-\varphi^*(\xi_i)]=0\).
These are obviously fulfilled when the conditional mean
\(\E[\varphi(\xi_i)\mid X_i,Z_i]\) is correctly
modelled. \correction{}{To address this both the stage 1 and stage 2
  model fits should be assessed using standard model checking tools
  for linear SEMs \citep{sanchez2009residual}.} Here the critical
distribution is that for the residual \(\widetilde{\zeta_i}\) of the
latent exposure, whereas correct specification of error terms in the
measurement model of the stage one model is less important as shown in
simulations of Section \ref{sec:sim}.  In Section \ref{sec:mix}, we
describe an extended method which allow flexible models for the
distribution of the latent exposure variable.



A few consequences of the calculations in the proof of Theorem \ref{thm:consistency}
 are important to note.
Firstly, it is important that the step-two model uses an unstructured
covariance matrix (\(\Psi_2\)) for the variance of the latent
variable. If that is not the case, 
 it may be impossible for the model to account for the
 misspecification of $\Sigma_{\eta}$ (\(\Sigma_{\eta} \ne
 \var(\eta_i)\)),
  therefore the modelled variance of \(V_i = (Z_i^t,\predphi(\xi_i)^t,Y_i^t)^t\)
may be wrong and estimation will likely
 become inconsistent. In
particular, the model should not be made identifiable 
by  fixing the variance of the latent variable to one. 
Secondly, covariates that are used in the step-two model, must also be
present in the first step. If that is not the case,  \(\cov[Z_i, \varphi(\xi_i)-
\predphi(\xi_i)]\) may not be zero and the step-two model will
have an incorrect variance leading to inconsistent estimation. Thirdly,
in the step-two model, the covariance terms \(\cov[Z,
\predphi(\xi_i)]\) and \(\var[\predphi(\xi_i)]\) are
modelled using unstructured matrices  so that any
information these terms might have had about the parameters is
disregarded in this approach. 
 \correction{
It is interesting to note that \(\cov[\predphi(\xi_i), \varphi(\xi_i)-
\predphi(\xi_i)]=0\), so that if \(\predphi(\xi_i)\)
is considered to be an imprecise version of \(\varphi(\xi_i)\) the
measurement error is of the Berkson-type \citep{carroll2006measurement}.}{} Finally, note that in Theorem \ref{thm:consistency}
it is assumed that variables in the stage one model affects the
variables in the stage two model only indirectly through the latent
variable. In the presence of a direct effect from one of the indicators,
\(X_i\), either on the latent variable or the indicators of the stage
two model, the proposed method can easily be extended by simply
including the relevant indicators as covariates in the stage two
model. The consistency of this approach is proven in
\appendixref{sec:app:consistent}. 

\correction{}{
It is interesting to compare our method to the two-stage
method-of-moments (2SMM) of 
\cite{wall2000estimation}. Here predictions \(\widehat{\xi}\) and \(\widehat{\eta}\) of the latent
variables are calculated from a confirmatory factor analysis model and
then the second step fits a non-linear 
regression model $\widehat{\eta}=
\beta \varphi(\widehat{\xi})+\zeta$ allowing for uncertainty in 
$\widehat{\xi}$. For the latter task a method moments estimator is
used, but it works only for polynomial models. Our method is
      different from 2SMM in two important ways. 
Rather than predicting \(\xi\) in step one,
we predict the non-linear terms \(\varphi(\xi)\) and therefore we are left
with a linear model in step two. Of course the predicted terms are different from the latent true
      terms, so, as in 2SMM, we have measurement error in
      covariates in the second step.  Here the choice of prediction
      method in step one
becomes important. We
  use the conditional mean \(\predphi(\xi_i)=\E[\varphi(\xi)\vert X,Z]\)  which has Berkson errors, that is the prediction in uncorrelated with the
      prediction error \(\cov[\predphi(\xi_i), \varphi(\xi_i)-
      \predphi(\xi_i)]=0\). In linear regression models,
      it is well-known that Berkson errors will not bias regression
      coefficients \citep{carroll2006measurement}. 
 In step two we estimate
parameters using a linear model
 and therefore the Berkson errors
      do not lead to inconsistency as we show in Theorem 1. In contrast,
      Bartlett predictions (used in 2SMM) have classic 
      error  and therefore adjustments
are needed in order to achieve unbiased estimation.}

\subsection{Extension to mixtures of structural equation models}
\label{sec:mix}

In this section we extend the structural part of the model to allow for
non-normal latent predictor variables. This is done through a mixture
model.  Thus, let \(G_i \sim \mbox{multinom}(\pi)\) be the class indicator
\(G_i\in\{1,\ldots,K\}\), and \(\xi_i\) the \(q\)-dimensional latent predictor
 \(\xi_i = \sum_{k=1}^{K}I(G_i=k)\xi_{ki},\)
where each component \(\xi_{ki}\)  follows a linear structural
equation
\begin{eqnarray}\label{mix1}
  \begin{array}{lcl} 
    \xi_{ki} & = & \tilde{\alpha}_k + \tilde{B} \xi_{ki}+\tilde{\Gamma} Z_i + \tilde{\zeta}_{ki},
  \end{array} 
\end{eqnarray}
with \(\tilde{\zeta}_{ki} \sim N(0, \tilde{\Psi}_k)\). We assume
\(G_i\) to be independent of
\((\tilde{\zeta}_{1i},...,\tilde{\zeta}_{Ki})\) and
\((\epsilon_i, \tilde{\epsilon}_i)\). Results can be extended also to the case where
\(\pi\) depends on covariates. Note, that the extension concerns only
the conditional distribution of the latent predictor given the
covariates. Thus, the only parameters that depend on \(k\) is the
intercept and the variance in the structural model for \(\xi\), whereas
the measurement models and the structural model for
\(\eta_i\) remain as shown in equations (\ref{stmodel}, \ref{memodely},
\ref{memodelx}) independent of \(k\).

In the extended model, the two-step procedure is modified by the fact
that ML-estimation in step 1 will be more complex (typically done via
the EM algorithm) and the predicted latent variables now become
\begin{eqnarray}\label{mixpredictions}
  \begin{array}{lcl} 
    \E[\varphi(\xi_i)\vert 
    X_i,Z_i]  &=& \E \{ \E [\varphi(\xi_i)\vert  X_i,Z_i,G_i] \vert X_i,Z_i \}\\
              &=& \sum_{k=1}^{K} P(G_i=k \vert X_i,Z_i) \, \E
                  [\varphi(\xi_{ki})\vert  X_i,Z_i,G_i=k]\\
              &=& \sum_{k=1}^{K} P(G_i=k \vert X_i,Z_i) \, \E
                  [\varphi(\xi_{ki})\vert  X_i,Z_i],
  \end{array} 
\end{eqnarray}
where the last step uses the fact that \(G_i\) is independent of
\((\zeta_{ki}, \tilde{\epsilon}_i)\). So the predictions of step 1 are
the sum of the product of the posterior probabilities
(class probabilities which are by-products of the EM algorithm) and
predictions of the type described in \techref{sec:prediction}. Of course, the second step in the estimation
procedure is unchanged: a linear SEM is fitted with the predictions
included as covariates.

\section{Prediction of non-linear latent terms} \label{sec:prediction}

In this section, we show  that for 
important classes of non-linear functions (\(\varphi\)) closed form expressions
can be derived for \(\predphi(\xi_i) = \E[\varphi(\xi_i) \vert X_i,Z_i]\).
Under the assumptions of the model,
 the conditional distribution of \(\xi_i\) given \(X_i\) and \(Z_i\) is normal with mean
and variance
\begin{eqnarray} \label{eogv}
  \begin{array}{cccccc} 
 m_{x,z} & =  & \E(\xi_i \vert X_i, Z_i)  & =  & \tilde{\alpha}+\tilde{\Gamma} Z_i + 
\Sigma_{X\xi}
\Sigma_X^{-1}
(X_i-\mu_X)\\

 v & = & \var(\xi_i \vert X_i, Z_i) & = & 
\tilde{\Psi} - \Sigma_{X\xi}
 \Sigma_X^{-1} \Sigma_{X\xi}^t,
\end{array} 
\end{eqnarray}

where \(\mu_X= \tilde{\nu}  +\tilde{\Lambda}(I-\tilde{B})^{-1} \alpha + \tilde{\Lambda}
(I-\tilde{B})^{-1} \tilde{\Gamma} Z_i + \tilde{K} Z_i, \Sigma_X= \tilde{\Lambda}
(I-\tilde{B})^{-1} \tilde{\Psi}(I-\tilde{B})^{-1t}\tilde{\Lambda}^{t}
+\tilde{\Omega}\) and \(\Sigma_{X\xi}=
\tilde{\Lambda}(I-\tilde{B})^{-1} \tilde{\Psi}(I-\tilde{B})^{-1t}\).
Note that the conditional variance does not depend on \(X\) and \(Z\). 
Below we will consider a number a non-linear models and provide
expressions for  \(\E[\varphi(\xi_i) \vert X_i,Z_i]\). In addition, we
will briefly compare our method to 
regression calibration \citep{carroll2006measurement}, where  the second
stage model simply would use \(\varphi(m_{x,z})\) as a
covariate in a linear SEM.

We start by considering univariate functions, so we assume  \(\xi_i\) to
be a scalar.
\begin{exa}
Polynomials 
\begin{eqnarray}\label{polynomial}
  \begin{array}{lcl} 
    \eta_i & = & \alpha + \sum_{m=1}^k \beta_m \xi_{i}^m+  
                 \zeta_i,
  \end{array} 
\end{eqnarray}

Here \(\varphi(\xi_i)=\xi_i^m \, (m \in \mathbb{N})\) and
conditional means are given by
\begin{eqnarray} \label{poly}
  \begin{array}{lcl} 
\E(\xi_i^m \vert X_i, Z_i)
=\sum_{k=0}^{[m/2]} \, m_{x,z}^{m-2k} \, v^{k} \, \frac{m!}{2^k k!
  (m-2k)!}
\end{array} 
\end{eqnarray}

In the quadratic model (\(\eta_i  =  \alpha + \beta_1 \xi_{i} + \beta_2
\xi_{i}^2 + \zeta_i\)), \(\E(\xi_i^2 \vert X_i, Z_i)=m_{x,z}^2+v\)
and therefore \(\E(\eta_i\vert X_i, Z_i)  =  \alpha + 
\beta_1 m_{x,z} + \beta_2 m_{x,z}^2 + \beta_2 v\). So in this 
model, regression calibration uses a correct expression for then mean
of \(\eta_i\) except that the term depending on \(v\) will not be included
and therefore the intercept will be estimated with bias. 
However, in a third-degree model,
\(\E(\xi_i^3 \vert X_i, Z_i)=m_{x,z}^3+3m_{x,y}v\) and
the regression calibration approach of just replacing \(\xi_i\)
with \(m_{x,z}\) will lead to biased coefficients of the polynomial.\\ 
\end{exa}

\begin{exa}
For the exponential function \(\varphi(\xi_i) =\exp(\xi_i)\) an
expression can be obtained as \(\exp(\xi_i)\) will follow a logarithmic
normal distribution where the mean is
\begin{eqnarray} \label{exp}
  \begin{array}{lcl} 
    \E[\exp(\xi_i) \vert X_i, Z_i]= \exp(0.5  v + m_{x,y})
  \end{array} 
\end{eqnarray}

Again, regression calibration is biased as the outcome
is regressed on \(\exp(m_{x,y})\) and not \(\exp(0.5
v+m_{x,y})\). If the conditional variance (\(v\)) is small
then the bias is expected also to be small.  The conditional mean of
functions on the form \(\varphi(\xi_i)=\exp(\xi_i)^m\) is
straightforward to calculate as this variable again follows a
logarithmic normal distribution. \\
\end{exa}

\begin{exa}
A piece-wise linear relation is described by 
\begin{eqnarray}\label{piecewiselin}
  \begin{array}{lcl} 
    \eta_i & = & \alpha  + \beta_1 (\xi_i 1_{\{ \xi_i <
  \tau\}} + \tau 1_{\{\xi_i > \tau\}} )+ \beta_2 ( \xi_i-\tau) 1_{\{ \xi_i >
  \tau\}} +
\zeta_i,
  \end{array} 
\end{eqnarray}
where \(\tau\) is a known break-point. As illustrated in 
\appendixref{sec:app:integralform} closed form expressions for the predictions
can be calculated, i.e.,
\begin{eqnarray} \label{special}
  \begin{array}{lcl} 
    \E(\xi_i 1_{\{\xi_i < \tau\}}\vert X_i, Z_i)& = & m_{x,z}
    \Phi(\frac{\tau-m_{x,z}}{s}) - s
    \phi(\frac{\tau-m_{x,z}}{s})\\

    \E[(\xi_i-\tau) 1_{\{\xi_i > \tau\}}\vert X_i, Z_i]
& = & (m_{x,z}-\tau) [1-\Phi(\frac{\tau-m_{x,z}}{s})] + s \phi(\frac{\tau-m_{x,z}}{s}),
  \end{array} 
\end{eqnarray}

where \(\phi\) and \(\Phi\) are the density function and the cumulative
distribution function of the standard normal distribution and
\(s=\sqrt{v}\). Clearly, 
regression calibration will generally provide 
inconsistent estimates in this model.

\end{exa}

\begin{exa}
A natural cubic spline with \(k\) knots \(t_1<t_2<\ldots <t_k\) is given by
\begin{eqnarray}\label{ns}
  \begin{array}{lcl} 
    \eta_i & = & \alpha + \beta_0 \xi_{i} + 
                 \sum_{j=1}^{k-2} \beta_j f_j(\xi_{i})+ 
                 \zeta_i,
  \end{array} 
\end{eqnarray}

with \(f_j(\xi_{i})=g_j(\xi_{i}) - \frac{t_k - t_j}{t_k-t_{k-1}} g_{k-1}(\xi_{i}) +
\frac{t_{k-1} - t_j}{t_k-t_{k-1}} g_{k}(\xi_{i}), \, j=1,\ldots ,k-2\) and
\(g_j(\xi_{i})=(\xi_{i}-t_j)^3 1_{\xi_{i}>t_j},  \, j=1,\ldots ,k\) \citep{durrleman1989flexible}. Thus, predictions 
\(\E[f_j(\xi_{i}) \vert X_i, Z_i]\)
are linear functions of \(\E[g_j(\xi_{i}) \vert X_i, Z_i]\).
In the \appendixref{sec:app:integralform} we derive the following
expressions for these means, i.e.,
\begin{eqnarray}\label{ns1}
  \begin{array}{lcl} 
\E[g_j(\xi_{i})\vert X_i, Z_i]
& = &\frac{s}{\sqrt{2 \pi}}[(2s^2+ (m_{x,z}-t_j)^2) 
\exp(-[\frac{(m_{x,z}-t_j)}{s\sqrt{2}}]^2)] + \\
& & (m_{x,z}-t_j) [(m_{x,z}-t_j)^2+3 s^2] p_{x,z,j},
  \end{array} 
\end{eqnarray}
where  \(s=\sqrt{v}\) and 
\(p_{x,z,j}=P(\xi_{i}>t_j \vert X_i, Z_i)=1-\Phi( \frac{t_j-m_{x,z}}{s})\).
Also, in this model, 
regression calibration will generally provide 
inconsistent estimates.

\end{exa}

The previous examples have focused on regression equations  with only
one latent predictor. Of course these calculations can easily be
extended to models with multiple predictors with non-linear effects as
long as the terms enter the model additively. Non-linear terms
depending on multiple latent variables are more complex. The last
example  describes 
the most common non-linear function involving two variables where
interactions are modelled using product terms.

\begin{exa}
Product-interaction model
\begin{eqnarray}\label{interaction}
  \begin{array}{lcl} 
    \eta_i & = & \alpha + \beta_1 \xi_{1i} + \beta_2 \xi_{2i} + \beta_{12} \xi_{1i}
                 \xi_{2i} + 
                 \zeta_i,
  \end{array} 
\end{eqnarray}

Now \(\E(\xi_{1i} \xi_{2i} \vert X_i, Z_i)=\cov(\xi_{1i}, \xi_{2i} \vert
X_i,Z_i)
+\E(\xi_{1i} \vert  X_i,Z_i) \E(\xi_{2i} \vert  X_i, Z_i)\), where
terms on the right-hand side are directly available
from  the bivariate
normal distribution of \(\xi_{1i}, \xi_{2i}\) given
\(X_i,Z_i\). Regression calibration leads to the correct mean expect
that the intercept will be biased as this method will not include the 
covariance term above.
\end{exa}

\section{Asymptotic properties of the two-stage estimator}
\label{sec:asym}

From Theorem \ref{thm:consistency} we have consistency of all
structural parameters in the stage 2 model. In this section, we show
that the limiting distribution of the estimator is asymptotically
normal, and we derive the asymptotic variance via the estimated
influence functions of the stage 1 and stage 2 estimators.

In the following we assume that the observations
\((Y_i, X_i, Z_i), i=1,\ldots,n\) are i.i.d., and we also restrict 
attention only to the consistent estimates as mentioned in the
previous section, i.e., \(\theta_2\) does not contain any of the parameters
belonging to \(\Psi\).  We will assume that the stage 1 model estimator
is obtained as the solution to the following score equation:
\begin{align}\label{eq:U1}
\mathcal{U}_1(\theta_1; X, Z) = \sum_{i=1}^n U_{1}(\theta_1; X_i, Z_i) = 0,
\end{align}
which typically will be the score of the usual Maximum Likelihood
Estimator. Similarly, if we could observe the latent variables in the
stage 1 model, a consistent estimator of the stage 2 model would be
obtained by solving a score equation corresponding to another linear
SEM
\begin{align*}
  \mathcal{U}_2(\theta_2; Y, X, Z, \xi) = \sum_{i=1}^n U_{2}\left(\theta_2; Y_i, Z_i, \phi(\xi_i)\right) = 0.
\end{align*}
As shown in Theorem \ref{thm:consistency}, a consistent estimator for
\(\theta_2\) can be obtained by instead considering the score equation
\begin{align}\label{eq:U2}
  \mathcal{U}_2(\theta_2; Y, X, Z) = \sum_{i=1}^n U_{2}\left(\theta_2; Y_i, Z_i, \predphi(X_i,Z_i)\right) = 0,
\end{align}
where \(\predphi(X_i,Z_i)=\E_{\theta_{01}}(\varphi(\xi_i)\mid X_i,Z_i)\).
Denote the simultaneous score function
\begin{align}\label{eq:U}
\mathcal{U}(\theta_2, \theta_1; Y, X, Z) = \sum_{i=1}^n U_{2}\left(\theta_2; Y_i, Z_i,
  \widehat\predphi(X_i,Z_i; \theta_1)\right),
\end{align}
where we have plugged in the predictions from the first model
\(\widehat{\predphi}(X_i,Z_i; \theta_1)=\E_{\theta_{1}}(\varphi(\xi_i)\mid X_i,Z_i)\) evaluated at the
parameter \(\theta_1\).

Let in the following \(\nabla_{\theta} \mathcal{U}(\theta)\) denote the \(m\times m\) matrix of partial
derivatives of \(\mathcal{U}\), where \(m\) is the dimension of the joint
parameter vector \(\theta\). We will impose the following regularity
conditions in addition to the consistency assumptions in Theorem
\ref{thm:consistency}:
\begin{enumerate}[label=(\alph*)]
\item The estimator of the stage 1 model is consistent, linear, regular, and asymptotic
  normal.\label{asump:stage1}
\item\label{asump:u2a} \(\mathcal{U}\) is twice continuous differentiable in a
  neighbourhood around the true (limiting) parameters
  \((\theta_{01}^T,\theta_{02}^T)^T\).  
 Further,
  \[
    n^{-1}\sum_{i=1}^n \nabla U_2(Y_i,X_i,Z_i; \theta_1, \theta_2) 
  \]
  converges uniformly to \(\E[ \nabla U_2(Y_i,X_i,Z_i; \theta_1, \theta_2)]\) in a neighbourhood around
  \((\theta_{01}^T,\theta_{02}^T)^T\),
\item and when evaluated here \(-\E(\nabla U_2)\) is positive definite
  \label{asump:u2b}
\end{enumerate}

We first note that assumption \ref{asump:stage1} means that
\begin{align*}
  \sqrt{n}(\widehat{\theta}_{1}-\theta_{01}) = \frac{1}{\sqrt{n}}\sum_{i=1}^{n}\influ_1(X_{i},Z_{i}; \theta_{01}) + o_{p}(1),
\end{align*}  
where \(\influ_1\) is the \emph{influence function} of the estimator \citep{stefanski_boos_2002_m-estimator}.
Similarly, (\ref{eq:U}) defines a regular consistent \(m\)-estimator,
and for known \(\theta_{01}\), it follows along the lines of
\citep[Theorem 3.4]{neweymcfadden1994}, \citep[Chapter 3]{tsiatis2006semiparametric}
from assumptions \ref{asump:u2a}-\ref{asump:u2b} that
\begin{align*} 
  \sqrt{n}(\widehat{\theta}_{2}(\theta_{01})-\theta_{02}) =
  \frac{1}{\sqrt{n}}\sum_{i=1}^{n}\influ_2(Y_i, Z_i, \predphi(X_i,Z_i); \theta_{02}) + o_{p}(1),
\end{align*}
where the influence function is given by
\begin{gather}
  \begin{split}
  &\influ_2(Y_i,Z_i,  \predphi(X_i,Z_i); \theta_{02}) = \\
  &\qquad 
  \E\left((-\nabla_{\theta_2}U_2(Y,Z,\predphi(X,Z),\theta_{02})\right)
  U_2(Y_i,Z_i,\predphi(X_i,Z_i),\theta_{02}).
  \end{split}
\end{gather}
A sufficient criterion for the uniform convergence is that the covariates are all bounded.

Denote \(\widehat{\theta_2}(\theta_1)\) as the estimator obtained from
solving (\ref{eq:U}) with fixed stage 1 parameter at \(\theta_1\). 
A Taylor expansion around \(\theta_{01}\) (see 
\citep[Chapter 6]{neweymcfadden1994}) yields
\begin{align*}
  \sqrt{n}(\widehat{\theta}_2(\widehat{\theta}_{1}) - \theta_{02}) =
  \sqrt{n}(\widehat{\theta}_2(\theta_{01}) - \theta_{02}) +
  \sqrt{n}\nabla\widehat{\theta}_2(\theta_{01})(\widehat{\theta}_{1}-\theta_{01}) + o_p(1).  
\end{align*}
This implies the following i.i.d. decomposition of the two-stage
estimator 
\begin{align}
  \begin{split}\label{eq:IF3}
    \sqrt{n}(\widehat{\theta_2}(\widehat{\theta_1}) - \theta_2)
    &= n^{-1/2}\sum_{i=1}^n \influ_2(Y_i,X_i,Z_i; \theta_2)  \\
    &\hspace*{-6ex}+ n^{-1/2}\,\E[-\nabla_{\theta_2} U_2(Y,Z,X; \theta_2,\theta_1)]^{-1}
    \\
    &\hspace*{-2ex}\times  \E[\nabla_{\theta_1}U(Y,Z,X; \theta_2,\theta_1)]
    \sum_{i=1}^n\influ_1(X_i,Z_i; \theta_1) + o_p(1) \\
    &= 
    n^{-1/2}\sum_{i=1}^n \influ_3(Y_i,X_i,Z_i; \theta_2,\theta_1) + o_p(1).
  \end{split}
\end{align}
By the central limit theorem we obtain that
\begin{align*}
  \sqrt{n}(\widehat{\theta_2}(\widehat{\theta_1}) - \theta_2)
  \overset{\mathcal{D}}{\longrightarrow} \mathcal{N}(0,\Sigma),
\end{align*}
where we can estimate the asymptotic variance by the plug-in estimate
\begin{align*}
  \widehat{\Sigma} = 
  \frac{1}{n}\sum_{i=1}^n \influ_3(Y_i,X_i,Z_i;
  \widehat{\theta}_2,\widehat{\theta}_1)^{\otimes 2}.
\end{align*}


As noted in \citep{parke_psedo_likelihood_estimation_annals1986}, the
two influence functions \(\influ_1\) and \(\influ_2\) will generally be
independent. However, we do not need to exploit this to obtain
consistent estimates of the variance.\\

The case where the stage one model is based on a Gaussian mixture
model requires additional considerations. Under the regularity
conditions specified in \citep{redner1984mixture}, we obtain the usual
rate of convergence such that
\(\sqrt{n}(\widehat{\theta}_1-\theta_1)\) is asymptotically normal distributed
with mean zero. This holds in particular for the important case, where
the conditional distribution in each group is Gaussian and the
conditional variance given covariates is fixed across the different
groups in the mixture model, i.e., only the intercept of \(\xi_{ki}\),
\((\tilde{\alpha}_k)\), varies between groups. It follows that in this case,
the necessary regularity conditions for the stage 1 model are
full-filled and the derivation of the asymptotic distribution of the
proposed two-stage estimator follows along the same lines as in the
normal case \(K=1\), with the influence function being estimated by
the product of the inverse information matrix and score function of
the MLE.
 


\section{Simulation study}
\label{sec:sim}

In this section we explore the finite sample properties of \correction{the
proposed two-stage procedure}{\twostage}.  We \correction{}{first} considered scenarios where all
distributional assumptions of the model were fulfilled\correction{
  but}{.
}
 Next, we explored the robustness of the proposed methods in misspecified
 models and finally we explored \twostage in a 
non-parametric
  setting using a flexible spline model. 
 The Monte Carlo simulations were based on the model illustrated in
Figure \ref{fig:path1} with the stage 1 \correction{}{model} defined as
\correction{\(\xi_i = -Z_i + \widetilde{\zeta}_i\)}
{\(\xi_i = \gamma_1Z_i + \widetilde{\zeta}_i\)}, and 
\correction{}{\(X_{ij} = \xi_i + \widetilde{\epsilon}_{ij}\)}
and the stage 2 model given by
\correction{}{
\begin{align*}
\eta_i = \beta^T\varphi(\xi_i) + \gamma_2Z_i + \zeta_i, \quad
Y_{ij} = \eta_i + \epsilon_{ij}, 
\end{align*}
} \correction{}{with mutually independent residual terms
  \(\widetilde{\zeta}_i, \zeta_i, \widetilde{\epsilon}_{ij}, \epsilon_{ij},
  j=1,2,3\). The distribution of these terms was varied throughout
  the simulations.}  The parameters of primary interest were the
structural parameters\correction{}{, \(\beta\), } defining the
association between the two latent variables \(\xi_i\) and \(\eta_i\).



\correction{Simulations I: Consistency}
{\subsection{Simulations I: Correctly specified model}}

\correction{We first conducted a simulation study to explore
 the properties of our estimator in correctly
 specified models.}{} Data was generated from a quadratic structural model
 \correction{
   \(\eta = \beta_{1}\xi + \beta_{2}\xi^{2} + Z + \zeta, \,
   \zeta \sim\mathcal{N}(0,1)\)}{
   \(\eta_i = \beta_{0} + \beta_{1}\xi_i + \beta_{2}\xi_i^{2} + \gamma_2Z_i + \zeta_i, \,
\)}
 with all residuals \(\widetilde{\zeta}_i, \zeta_i, \widetilde{\epsilon}_{ij}, \epsilon_{ij},
  j=1,2,3\) being standard normal and
 \(\beta_{0}=1, \beta_{1}=1, \beta_{2}=0.5\). The simulation was run with sample sizes of
\correction{\(n=500\)}{\(n=1,000\) and \(n=500\) (see
  \techref{appendix:simulation})}{, without any covariate in the model
  (\(\gamma_1=0, \gamma_2=0\))  and with a covariate
(\(\gamma_1= 1, \gamma_2=1\), see supplementary material)}. The
simulation study was based on \correction{500}{1,000}
replications. The \twostage methods were compared to Bollen's
two-stage least squares (2SLS) estimator \citep{bollen1995structural}
(see \techref{sec:bolleniv}), 2SMM of
\cite{wall2000estimation}, and approximate ML based on a Laplace
approximation as well as Adaptive
Gaussian Quadrature (AGQ) with 9 quadrature points.
 
\begin{table}
  \centering
\begin{tabular}[t]{llllllll}
  \toprule
  \bfseries{} & \bfseries{} & \bfseries{Mean} & \bfseries{SD} & 
  \bfseries{SE} & \(\tfrac{\text{\bfseries{SE}}}{\text{\bfseries{SD}}}\) & \bfseries{Cov.} & \bfseries{RMSE}\\
  \midrule
   & \twostage & 0.996 & 0.371 & 0.349 & 0.942 & 0.940 & 0.371\\

 & \twostage mixture & 1.031 & 0.393 & 0.424 & 1.079 & 0.941 & 0.394\\

 & 2SLS & 0.975 & 0.647 & 0.606 & 0.938 & 0.938 & 0.647\\

 & 2SLS robust & 0.975 & 0.647 & 0.631 & 0.975 & 0.948 & 0.647\\

 & 2SMM & 1.031 & 0.431 & {} & {} & {} & 0.432\\

 & 2SMM robust & 1.042 & 0.466 & {} & {} & {} & 0.468\\

 & Laplace & 1.127 & 0.306 & 0.292 & 0.957 & 0.939 & 0.331\\

  \multirow{-8}{*}{\raggedright\arraybackslash \(\bm{\beta_1=1}\)}
              & AGQ9 & 1.002 & 0.277 & 0.272 & 0.984 & 0.947 & 0.277\\
  \midrule
 & \twostage & 0.499 & 0.072 & 0.068 & 0.944 & 0.933 & 0.072\\

 & \twostage mixture & 0.507 & 0.077 & 0.084 & 1.095 & 0.934 & 0.077\\

 & 2SLS & 0.496 & 0.115 & 0.100 & 0.864 & 0.919 & 0.115\\

 & 2SLS robust & 0.496 & 0.115 & 0.112 & 0.973 & 0.942 & 0.115\\

 & 2SMM & 0.506 & 0.081 & {} & {} & {} & 0.081\\

 & 2SMM robust & 0.508 & 0.088 & {} & {} & {} & 0.088\\

 & Laplace & 0.526 & 0.060 & 0.057 & 0.958 & 0.939 & 0.065\\

  \multirow{-8}{*}{\raggedright\arraybackslash \(\bm{\beta_2=0.5}\)}
              & AGQ9 & 0.501 & 0.054 & 0.053 & 0.986 & 0.943 & 0.054\\
  \bottomrule
\end{tabular}
\caption{\label{sim1} Performance of the estimators: Gaussian \twostage,
  2SEMM with 2-component mixture (\twostage mixture), 2SLS, 2SLS with
  heteroskedasticity standard errors (2SLS robust), methods of moments
  estimator (2SMM and 2SMM robust, with the former
  deriving moments from a Gaussian distribution. Standard error
  are omitted here but are derived in \cite{wall2000estimation}), and approximate ML
  (Laplace, AGQ9) in a simulation study from a quadratic model
  \(\E(\eta\mid\xi) = \beta_0 + \beta_1\xi + \beta_2\xi^2\)
  (true parameters \(\beta_1=1\) and \(\beta_2=0.5\)) where all assumptions hold.}
\end{table}

Simulations showed that the \twostage estimator had good properties in
finite samples (Table \ref{sim1}).  The method seemed approximately
unbiased and confidence intervals had coverage probabilities close to
the nominal level. It is interesting to see that in this case, where
residual terms were normal, nothing seemed to have been lost by
applying the robust mixture model extension. In addition to providing
\correction{}{effectively} unbiased inference this method yielded
standard errors that were very close to those obtained assuming
normality.  As expected, ML analysis was more efficient, but the loss
of the \twostage procedure was modest. The Laplace approximation
showed some bias compared to the rest of the methods, but preformed
well as measured by the RMSE and was almost as efficient as the more
sophisticated AGQ-approximation.  Bollen's 2SLS \correction{method}{}
provided unbiased estimation, but it was clearly less efficient than
both ML and \twostage \correction{estimation}{}. Also, the non-robust
standard errors as suggested by \cite{bollen1995structural}
underestimated the uncertainty in the second order
term. \correction{}{The methods of moments estimator 2SMM was less
  efficient than the \twostage estimator but performed clearly better
  than \correction{the}{} 2SLS\correction{estimator}{}. The
  conclusions were consistent across all scenarios we examined (see
  \techref{appendix:simulation}).}

\correction{}{
  In contrast to 2SLS and 2SMM, our method is not restricted to
  polynomial structures, and we also
  examined the performance with an exponential effect, \(\eta_i = \beta_{1}\xi_i + \beta_2\exp(\xi_i) +
  \zeta_i\). The \twostage estimator also performed well in this
  setting being effectively unbiased with correct coverage of
  the confidence limits (see \techref{appendix:simulation} for
  details). 
}

\correction{
We also assessed the performance of the model with an exponential effect
\(\eta = \beta_{1}\xi + \beta_2\exp(\xi) + \zeta.\)
The consistency of the
2SLS estimator relies on the assumption that the function \(\varphi\) can be
decomposed additively as
\(\varphi(\xi) = \varphi(x_1-\widetilde{\epsilon}_1) = g_1(x_1) + g_2(x_1,\widetilde{\epsilon}_1)\) and with access to
instrumental variables (indicators \(X_i\)) that should be
uncorrelated with the second term, \(g_2\). This is not possible
with the exponential function, however in a neighbourhood around zero 
\(\exp(x_1-\widetilde{\epsilon}_1) \approx \exp(x_1) + \exp(x_1)(x_1-\widetilde{\epsilon}_1)\), which suggests a
first order approximate 2SLS estimator with  \(g_1(x_1) = \exp(x_1)\).
}{}

\correction{
The results of the simulation with an exponential transformation are
summarized in Table \ref{sim1}. As expected we see that the
two-stage estimator is unbiased while this is no longer the case for
the 2SLS estimator. Also, the coverage of the two-stage estimator
remains close to the nominal level, however with more extreme choices
of the parameters (simulations not shown) we did observe that the
sample size needed to be increased to obtain correct coverage levels.}{}

\subsection{Simulations II: Robustness}

\begin{table}[!h]
  \centering
  \begin{tabular}[t]{llllllll}    
    \toprule    
    \bfseries{} & \bfseries{} & \bfseries{Mean} & \bfseries{SD} & \bfseries{SE} & \bfseries{\(\tfrac{\text{\bfseries{SE}}}{\text{\bfseries{SD}}}\)} & \bfseries{Cov.} & \bfseries{RMSE}\\
    \midrule
    \(\widetilde{\zeta}\sim GMM\) \\
    \midrule    
 & \twostage & 1.349 & 0.088 & 0.086 & 0.976 & 0.021 & 0.360\\

 & \twostage mixture  & 1.000 & 0.104 & 0.103 & 0.988 & 0.948 & 0.104\\

 & 2SLS robust & 1.012 & 0.169 & 0.162 & 0.956 & 0.936 & 0.170\\

 & 2SMM & 1.000 & 0.112 & {} & {} & {} & 0.112\\

    \multirow{-5}{*}{\raggedright\arraybackslash \(\bm{\beta_1=1}\)}
                & 2SMM robust & 0.999 & 0.114 & {} & {} & {} & 0.114\\
    \midrule
 & \twostage & 0.380 & 0.030 & 0.029 & 0.976 & 0.025 & 0.123\\

 & \twostage mixture & 0.499 & 0.038 & 0.038 & 0.990 & 0.947 & 0.038\\

 & 2SLS robust & 0.496 & 0.056 & 0.055 & 0.976 & 0.929 & 0.056\\
    
 & 2SMM & 0.500 & 0.040 & {} & {} & {} & 0.040\\

    \multirow{-5}{*}{\raggedright\arraybackslash \(\bm{\beta_2=0.5}\)}
                & 2SMM robust & 0.500 & 0.041 & {} & {} & {} & 0.041\\
    
    \midrule 
    \(\widetilde{\zeta}\sim Unif\) \\
    \midrule
 & \twostage & 0.998 & 0.069 & 0.069 & 0.998 & 0.948 & 0.069\\

 & \twostage mixture & 0.998 & 0.075 & 0.075 & 1.007 & 0.957 & 0.075\\


 & 2SLS robust & 1.003 & 0.085 & 0.087 & 1.015 & 0.959 & 0.086\\

 & 2SMM & 1.001 & 0.078 & {} & {} & {} & 0.078\\

    \multirow{-5}{*}{\raggedright\arraybackslash \(\bm{\beta_1=1}\)}
                & 2SMM robust & 1.001 & 0.077 & {} & {} & {} & 0.077\\
    
    \midrule
 & \twostage & 0.310 & 0.054 & 0.056 & 1.030 & 0.102 & 0.198\\

 & \twostage mixture & 0.485 & 0.101 & 0.104 & 1.027 & 0.944 & 0.102\\


 & 2SLS robust & 0.489 & 0.160 & 0.159 & 0.994 & 0.934 & 0.161\\

 & 2SMM & 0.544 & 0.124 & {} & {} & {} & 0.131\\

    \multirow{-5}{*}{\raggedright\arraybackslash \(\bm{\beta_1=0.5}\)}
                & 2SMM robust & 0.520 & 0.126 & {} & {} & {} & 0.128\\
    
    \midrule
    \(\widetilde{\epsilon}_j\sim Unif\) \\
    \midrule
 & \twostage & 0.997 & 0.075 & 0.075 & 1.002 & 0.945 & 0.075\\

 & \twostage mixture & 0.998 & 0.082 & 0.082 & 1.000 & 0.935 & 0.082\\


 & 2SLS robust & 0.997 & 0.090 & 0.091 & 1.014 & 0.951 & 0.090\\

 & 2SMM & 1.000 & 0.079 & {} & {} & {} & 0.079\\

    \multirow{-5}{*}{\raggedright\arraybackslash \(\bm{\beta_1=1}\)}
                & 2SMM robust & 1.000 & 0.079 & {} & {} & {} & 0.079\\
    \midrule

 & \twostage & 0.506 & 0.063 & 0.063 & 0.996 & 0.951 & 0.063\\

 & \twostage mixture & 0.521 & 0.072 & 0.076 & 1.055 & 0.956 & 0.075\\


 & 2SLS robust & 0.499 & 0.088 & 0.088 & 0.996 & 0.937 & 0.088\\

 & 2SMM & 0.521 & 0.070 & {} & {} & {} & 0.073\\

    \multirow{-5}{*}{\raggedright\arraybackslash \(\bm{\beta_1=0.5}\)}
                & 2SMM robust & 0.512 & 0.074 & {} & {} & {} & 0.074\\

    \bottomrule    
  \end{tabular}
  \caption{Performance of the two-stage estimator assuming a Gaussian
  distribution (\twostage) and with  2-component mixture (\twostage mixture) in a
  simulation study from a quadratic model in three scenarios,
where modelling assumptions
  were not all satisfied. First the latent predictor
  followed a two component mixture distribution, then the latent predictor
  followed a uniform distribution and finally
  the residuals in the measurement model of stage one followed a uniform distribution.\label{sim2}}
\end{table}

Here we explored the properties of the estimators in misspecified
models.  First we examined data generating mechanisms \correction{where}{identical to the
previous model except of} the
conditional distribution of the latent variable \(\xi_i\) \correction{}{which} was not Gaussian
but followed a mixture distribution,
\correction{
\(\widetilde{\zeta} \sim
GMM(p_{1}=0.25,\mu_{1}=0,\mu_{2}=3,\sigma^{2}=1)\).
}{i.e., \(\widetilde{\zeta}_i \sim 0.25\mathcal{N}(0,1) + 0.75\mathcal{N}(3,1)\).}
The results are
summarized in Table \ref{sim2}, where we see some bias in the
\twostage estimator using a Gaussian distribution for the stage 1
model. The mixture \twostage estimator is unbiased with correct
coverage.  As an observation, we noted that \correction{our proposed two-stage
estimation procedure}{\twostage} was generally much faster, more computational
stable and less dependent on starting values than the Laplace and AGQ approximations. This was especially
the case in the mixture setting where the ML-methods had convergence
problems and need for fine-tuning across different implementations
(results not shown).

The above simulation setup corresponds exactly to the assumptions of
our mixture model extension, so to test the robustness of the
extension we also included a study where \(\widetilde{\zeta}_i\)
followed a uniform distribution with mean zero and variance one, and a
simulation where the residuals of the indicators,
\(\widetilde{\epsilon}_{ij}\) followed a uniform distribution. In both
cases, the \twostage mixture estimator was effectively unbiased with
coverage close to the nominal level. Interestingly, the
\correction{}{Gaussian \twostage} was robust to the misspecification
of the indicator distribution where it performed slightly better than
the mixture model.  \correction{}{Both 2SLS and 2SMM appeared to be
  robust to the considered misspecifications. In most cases both
  estimators were less efficient than \twostage (see
  \techref{appendix:simulation}), however, as the sample size
  increased we observed that 2SMM seemed to catch up with
  \twostage. We note however, that a severe limitation of the 2SMM
  approach is the lack of generalizations (and implementations)
  allowing for example relaxation of conditional independence
  assumptions, inclusion of covariates, and most importantly
  specifications of functional forms beyond the polynomial structure.}

\correction{}{
  \subsection{Simulations III: Non-parametric estimation}
 
  To study \twostage in a non-parametric setting, we also simulated
  data from the measurement models defined
  in the previous sections but with unknown functional relationship
  between the latent variables given by $\varphi(\xi; \bm{\beta}) = \beta_1\xi + \beta_2\xi^2 + \sin(\beta_3\xi)$.
A natural cubic spline with \(k\) knots \(t_1<t_2<\ldots <t_k\)
is given by $\E(\eta_i\mid \xi_i) =  \gamma_0 + \gamma_1 \xi_{i} + 
                 \sum_{j=1}^{k-2} \gamma_{j+1} f_j(\xi_{i})$
                 , with \(f_j(\xi_{i})=g_j(\xi_{i}) - \frac{t_k - t_j}{t_k-t_{k-1}} g_{k-1}(\xi_{i}) +
\frac{t_{k-1} - t_j}{t_k-t_{k-1}} g_{k}(\xi_{i}), \, j=1,\ldots
,k-2\). Here $g_j(\xi_{i})=(\xi_{i}-t_j)^3 1_{\{\xi_{i}>t_j\}}$ so to
apply \twostage  we calculated \(\E[g_j(\xi_{i}) \vert X_i, Z_i]\)
(see \techref{sec:prediction}).
  As a benchmark we compared \twostage with the estimator proposed by
  \cite{kelava2017} and the corresponding Matlab
  implementation\footnotemark\footnotetext{\url{https://github.com/tifasch/nonparametric/tree/ead709097d6}}.
  Here the number of equidistant knots were chosen by
  dividing the simulated data into a single test and training data
  of equal size and choosing the spline basis (degrees of freedom
  varying from 1 to 11) as the one that minimized the RMSE evaluated
  in the test data.  We noted that slightly better results were obtained for \twostage when the hyper-parameters (spline
  knots) were chosen using 5-fold cross validation.  To
  make the results more comparable we, however, adopted the same method  
  for choosing the degrees of freedom for the spline using the exact
  same split of the testing and training data.

  In each simulation, \(r=1,\ldots,100\), we generated \(n=200\)
  observations, and for each estimator we calculated
  \(
      \text{RMSE}_{r} = \big(\sum_{i=1}^n \left[\varphi(\xi_{i};
    \bm{\beta}) - \tilde\varphi(\xi_i; \widehat{\bm{\gamma}}_r)\right]^2\big)^{\tfrac{1}{2}},
\)
  where \(\bm{\beta}\) denotes the true parameter and
  \(\widehat{\bm{\gamma}}_{r}\) is the estimated parameters of the
  spline model, i.e.,
  \(\widehat{\eta} = \tilde\varphi(\xi; \widehat{\bm{\gamma}}_r) =
  \bm{B}(\xi)\widehat{\bm{\gamma}}_r\), where \(\bm{B}(\xi)\) is the
  spline basis design matrix.  With
  \(\beta_1=1, \beta_2=0, \beta_3=1\) and with
  \(\widetilde{\zeta}, \zeta\sim\mathcal{N}(0,1)\) the average RMSE
  over all replications were 0.314 for Kelava's estimator and 0.112
  for \twostage, and similarly when
  \(\widetilde{\zeta}, \zeta \sim U(-6,6)\) the average RMSE were
  0.933 and 0.608 in favour of \twostage.  Similar conclusions were
  drawn when using a stronger non-linear functional form given by
  \(\beta_1=1, \beta_2=0, \beta_3=1\). In the Gaussian case the
  average RMSE was 1.177 and 0.827, and in the uniform case 3.613 and
  1.988, all in favour of \twostage.  In addition, we note that
  another advantage of the \twostage estimator is the immediately
  available expressions of the asymptotic variance through the
  estimated influence functions. See \techref{sec:simsupnonpar} for
  more details.

}


\section{Application: Modelling in vivo brain serotonin measurements}

Serotonin (5-HT, 5-hydrotryptamine) is known to play an important role
in the regulation of appetite, sleep, mood, sex, and memory
function. Variation in cerebral 5-HT levels is also recognized as
 being influential on addiction and development of psychiatric
disorders such as schizophrenia and depression.  With Positron
Emission Tomography (PET) techniques it is possible to
quantify post and pre-synaptic markers of the serotonergic system in
the living human brain, such as the serotonin 2A receptor (\hta) and
the serotonin transporter (\htt). \correction{These markers have been intensively
studied and associations to various endpoints such as obesity and mood disorders
\citep{meyer07:_imagin,Meyer:1999uq} and obesity
\citep{erritzoe2009brain} have been identified.}{
These markers have been intensively
studied and associations to eating, sleeping, and mood disorders
\citep{meyer07:_imagin,Meyer:1999uq} have been identified.}

Animal studies examining the consequence of manipulation of central
5-HT levels indicate an approximate (negative) linear relationship
between normal synaptic 5-HT levels and \hta receptor binding
\correction{\citep{pmid17124620,licht09:_ht4_flind_sensit_line}}
{\citep{licht09:_ht4_flind_sensit_line}}. It has been
suggested that
\correction{\citep{meyer07:_imagin,haugboel07:_cereb_ht2a_touret}}
{\citep{meyer07:_imagin}}
\hta receptor binding may act as an indicator of the cerebral 5-HT
levels. Similarly, experimental studies have shown that manipulation
of synaptic 5-HT levels causes change in the \htt binding
\correction{\citep{gould06,pineyro94:_desen_ht}}
{\citep{pineyro94:_desen_ht}} with a suggested non-linear
functional form (inverted u-shape where low and high serotonin levels
both are associated with low levels of \htt). This association was
studied in \citep{DavidErritzoe03032010} hypothesizing that underlying
low 5-HT levels could lead to a compensatory up-regulation of \hta
receptor binding and down-regulation of \htt.

\begin{figure}[htbp]
  \centering
  \mbox{
    \raisebox{0.2\height}{\includegraphics[width=0.50\textwidth]{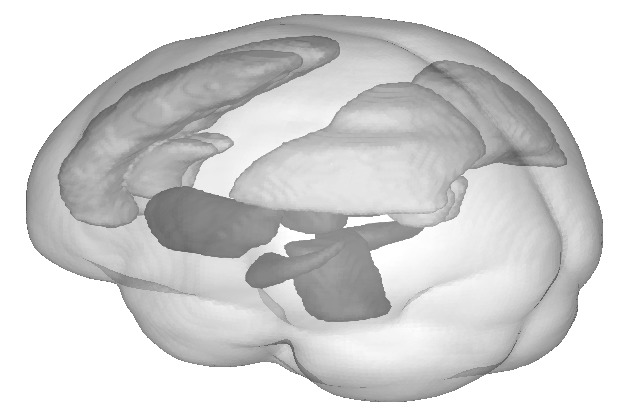}}
    \includegraphics[width=0.45\textwidth]{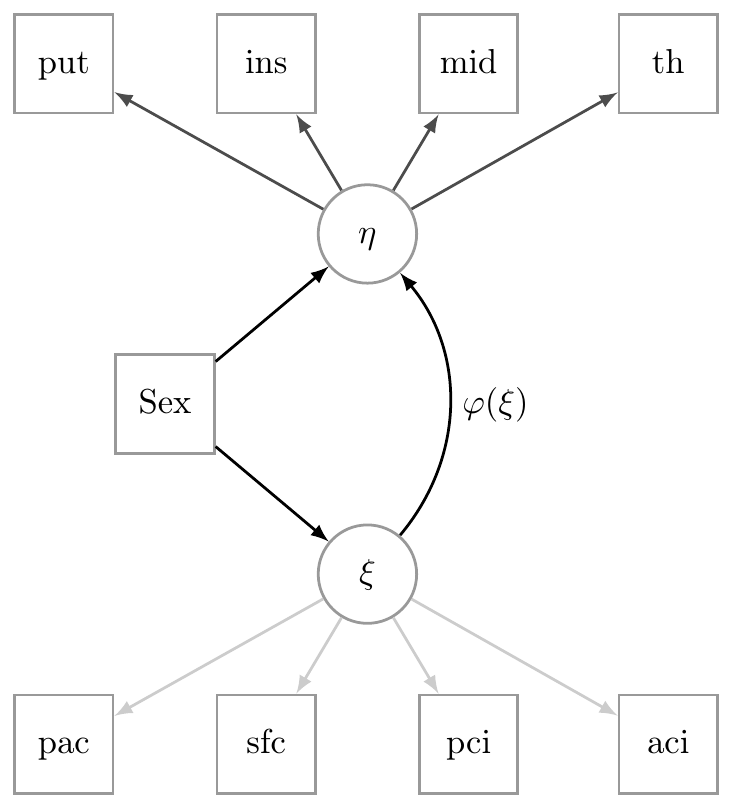}
  }
  \caption{\label{fig:2asem1}Path diagram for a structural equation model describing
    the relationship between \hta receptor binding (light grey) and \htt binding
    (dark grey). Each of the two types of markers are described by a single
    latent variable. For the \hta receptor the following regions
    (light grey regions in the glass brain of the right figure) was
    chosen as measurements:
     \emph{Parietal cortex} (pac), \emph{Superior
      frontal cortex} (sfc), \emph{Posterior cingulate gyrus}
    (pci), and \emph{Anterior cingulate gyrus} (aci). For the
    serotonin transporter we chose the regions: \emph{Putamen} (put),
    \emph{Insula} (ins), \emph{Midbrain} (mid), \emph{Thalamus}
    (th).}  
\end{figure}

We will here present an analysis of the same sample as in the original
paper, while taking into account the measurement error in
both the \hta and \htt measurements by using a non-linear \correction{structural
equation model}{SEM}.  \correction{}{The} \hta receptor binding
potential (\bpp) and \correction{}{} the \htt
binding potential (\bpnd) was measured in \correction{a sample of}{} 56 normal
subjects. For each subject, the measurements were summarized in a
number of regions of interest. We refer to the original paper for
details on the method used in acquiring the data.

For the serotonergic markers the concept of a measurement model seems to
be ideal in capturing the idea of an underlying common regulator of the
two types of measurements. For the \hta receptor 
outcome in a given region, we will assume a model
\begin{align}
  \bpp{}_{,\mathrm{ROI}} = \mu_{\mathrm{ROI}} + \lambda_{\mathrm{ROI}}\cdot\xi + \epsilon_{\mathrm{ROI}},
\end{align}
with a single latent variable \(\xi\).  The flexibility in letting
each region have its own intercept, \(\mu_{\mathrm{ROI}}\), loading parameter,
\(\lambda_{\mathrm{ROI}}\), and residual term
\(\epsilon_{\mathrm{ROI}}\sim\mathcal{N}(0,\sigma_{\mathrm{ROI}}^2)\) allows us to model
data, where different regions have different degrees of binding
potential, variation and correlation with other regions. We will
assume independence between residuals though this is not a necessary
assumption. We propose a similar model for \htt binding with a
measurement model described by a latent variable \(\eta\).
For both markers we chose 4 high binding regions of interest which
previously have been demonstrated to be reliable measurements of \hta
and \htt binding, respectively (see Figure \ref{fig:2asem1}).
To describe the association between \(\hta\) receptor binding and \htt
binding we added a simple structural model \(\eta = \mu_{sex} + \beta \xi + \zeta\),
to see how well a linear approximation would describe the
relationship.  In this simple model, \(\xi\) takes the role of the
common regulator, i.e. the central 5-HT level, as measured directly by
the \hta receptor binding, and the common regulator predicts the levels of
the global \htt variable \(\eta\).

We estimated the parameters of the model using ML\correction{-estimation}{}. The estimate of the primary parameter of interest,
\(\beta\), was 0.046 \bpnd/\bpp (with parietal cortex and thalamus as
reference regions, allowing us to interpret the effect as change in
SERT \bpp in thalamus per unit change in parietal cortical \bpnd) and
95\% CI \([-0.057; 0.149]\). The lack of statistical
significance may be explained by the lack of non-linear effects in our
model specification.  A \(\chi^2\) omnibus-test of goodness-of-fit (a
likelihood ratio test against an unstructured 8-dimensional normal
distribution) yielded a \(p\)-value of \(0.23\). Thus, based solely on this
test there was no evidence against the model. Clearly, this
goodness-of-fit test is, however, not adequate for detecting
non-linearities \correction{}{\citep{mooijaart2009insensitivity}}.

\begin{figure}[!htb]
\centering
\includegraphics[width=0.95\textwidth]{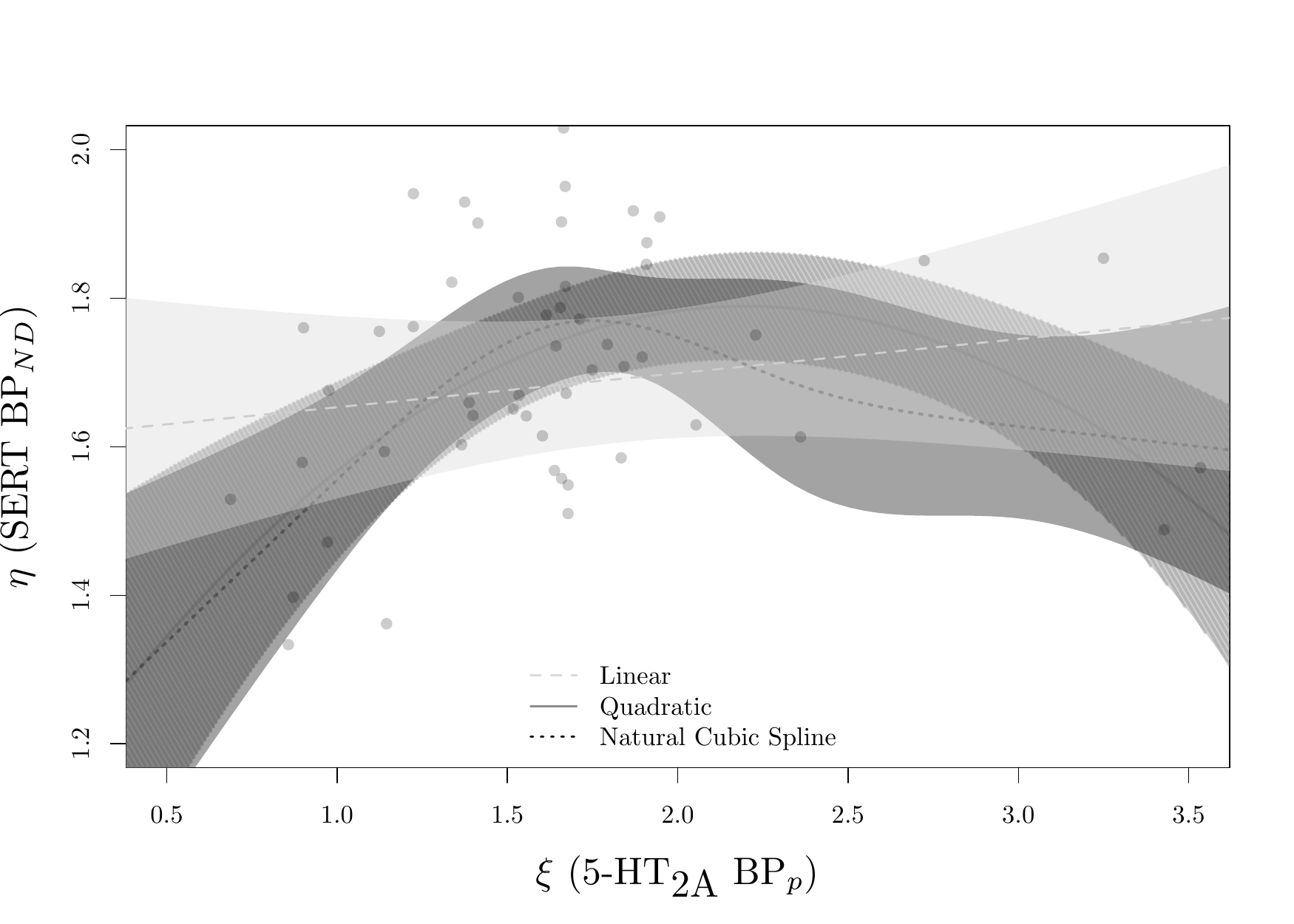}
\caption{Association between \hta \bpp binding potential and \htt
  \bpnd binding potential as estimated using linear, quadratic and
  natural cubic spline model. \label{fig:2asert1}.
  The points are the Empirical Bayes estimates from two separate
  linear SEMs.}
\end{figure}

\correction{Using the two-stage estimator}{Next, we applied the
  \twostage procedure to} the structural
equation model of Figure \ref{fig:2asem1} with the association between
\(\xi\) and \(\eta\) described by a second order polynomial: \(\eta = \mu_{sex} + \beta_1\xi + \beta_2\xi^2 + \zeta\).
The estimates were \(\widehat{\beta}_1=0.676\) (95\% CI \([0.321; 1.030]\), \(p=0.00018\)) and
\(\widehat{\beta}_2=-0.153\) (95\% CI \([-0.233;
-0.074]\), \(p=0.0002\)), thus confirming our hypothesis of a
non-linear association between \hta receptor and \htt binding
potential (Wald test for the hypothesis of no association:
\(p=0.0008, df=2\)).

A more flexible natural cubic spline model was next applied. The
predicted latent variable of the measurement error model for \hta
receptor binding potential was in the range 0.5-3.5 \bpp, and we
choose 4 knot points equidistantly \correction{distributed}{} in the interval 1 to 3.
The association between the two markers (and comparison with the
linear and quadratic model) is shown in Figure \ref{fig:2asert1}. The
natural cubic spline suggests a more flat association between \hta
\bpp{} and SERT \bpnd for high \hta binding potential values, but
otherwise there was a close agreement with the quadratic model.  We
conducted a more formal comparison of the two models using 5-fold
cross validation with all indicator variables normalized. The RMSE was
in slight favour of the quadratic model (0.94 vs 1.03). We also
examined natural cubic splines with increased number of knots, but
they all exhibited over-fitting with \correction{worse}{higher} RMSE. 
Finally, we used a two-component mixture in the stage
one model and got results that were
almost identical to those of the Gaussian
model. Also, as in the original paper we observed that the estimated
non-linear association was not sensitive to removing the observations
with the highest values of \hta binding potential from the data
(results not shown).

Thus, in agreement with animal and experimental studies, we were able
to show a non-linear association between these two serotonergic
markers. Our refined analysis also confirmed the findings of the
original paper \citep{DavidErritzoe03032010}, where the same data was
analysed using standard regression techniques and hence
\correction{is} results are likely to
be susceptible to bias due to measurement error in both variables.

\section{Discussion}

\correction{ML-inference in non-linear structural equation models
is complex. Computational intensive
methods based  on numerical integration are needed and expert knowledge    
is required to tweak algorithms in case of problems with
non-convergence. Furthermore,  violations of assumptions of normality
are difficult to separate from effects of a non-linear transformation,
therefore ML-inference in the non-linear model is not
robust to distributional assumptions. These challenges have surely
limited the number of applications of this interesting class of
models.}{ML-inference in non-linear SEMs
is complex. Computational intensive
methods based  on numerical integration are needed and results are
sensitive to distributional assumptions.}  
\correction{In this paper we presented a computational simple two-stage estimator
for non-linear models.}{This paper presented the two-stage estimator
\twostage as a computationally simple
alternative to ML.} Here both steps are based on linear models:
first we predict the non-linear terms and then these are related
to latent outcomes in the second step. We identified the asymptotic
distribution of \correction{our estimator}{\twostage}, developed a robust extension based on
mixture models and implemented the methods in a user-friendly
\texttt{R}-package (see \appendixref{appendix:software}). Simulations indicated
a modest loss of efficiency compared to ML-estimation and our method
was shown to be \correction{much more powerful than Bollen's IV, which is another
computationally simple and robust alternative to ML.}{
more powerful than two 
computationally simple and robust alternatives, i.e., 2SLS and 2SMM}.
In addition, \correction{to
providing superior power}{} \twostage can  be applied to a larger
class of non-linear functions than \correction{IV}{2SLS and 2SMM}. In particular the class of
restricted cubic splines is an important example which have shown to
be very useful in applications of regression models. The introduction
of stable and fast estimation algorithms for spline functions in the
structural equation framework is likely to lead to important
improvements in applications which for too long have been restricted
to linear relationships. 

In linear models, \correction{the two-stage approach}{\twostage} is
equivalent to regression calibration
\citep{carroll2006measurement}. This method has been investigated in
linear SEMs e.g. by \cite{skrondal2001regression} and
\cite{sanchez2009estimating}. In non-linear SEMs, the idea of using
mixture modelling to achieve more robust estimation has been exploited
for ML-estimators e.g. by \cite{kelava2014nonlinear}.  The handling of
splines in \correction{the two-stage estimator}{\twostage} is related
to the non-parametric estimators suggested by
\cite{carroll1999nonparametric} for linear regression models with
measurement error in covariates.  Bayesian methods have been developed
for semi-parametric estimation in SEMs (\cite{song2013bayesian},
\cite{guo2012bayesian}, \cite{kelava2014general}), but frequentist
methods are rare.  \cite{bauer2005semiparametric} and
\cite{kelava2017} presented interesting methods but did not provide
results on asymptotic standard errors. \correction{}{The latter
  procedure was included in our simulation study where it yielded
  larger prediction errors than \twostage.}

The two-stage approach may be especially useful in data bases with
many different research projects. Here the SEM
for the exposure may be fitted only once and the predicted non-linear
terms can be stored along with the influence function. Then the
predictions of exposure terms can be related to different outcomes 
by different research groups using linear structural models with
corrected standard errors.  Even in situations where ML-inference is
the goal, \correction{the two-stage approach}{\twostage}
will likely be very useful in providing good starting values.

\correction{We included IV-method
in our simulation study and our conclusion about this
  method is, that it is in general inferior
to our two-stage estimator.  IV is less powerful and in many
non-linear models it will lead to biased inference. However, for the
quadratic and the product-interaction model it is consistent also in
models with non-normal predictors. So in these models, IV will be a
useful supplement to ML and two-stage estimation. However, our
simulations clearly demonstrated that inference based on the usual IV
standard errors is invalid and should be avoided.}{

}

\correction{}{ When using \twostage for assessing associations between
  latent variables an obvious strategy would be to start the analysis
  with a rich parametric model (e.g., spline model) which may then be
  reduced by backward selection using Wald tests.
  An obvious extension would be to develop lasso-type regularization for the \twostage estimator
  Similarly, different parametric forms may be compared using Wald
  tests in a nested model. An alternative is to base the model
  selection on the estimated out-of-sample predictive performance
  through cross-validation as demonstrated in the application.  It may
  also be possible to develop fit criteria to detect general non-linear
  misfit. Recent theory in non-linear SEMs have focused on the
  development of such criteria and it may be possible to extend these
  so that they can be used together with \twostage. For example,
  \cite{schermelleh2014evaluation} developed a $\chi^2$-test comparing
  the observed and expected covariance matrix of the observed
  variables appended with selected products of indicators. Another
  interesting possibility would be to consider cumulative residuals as
  described by \cite{sanchez2009residual} for linear SEMs.}

We applied non-linear models
to PET measurements of the serotonergic system.
Based on biological evidence we proposed a statistical model
describing the underlying cerebral 5-HT level by inclusion of latent
components in a SEM. The underlying 5-HT level
were here assumed to be measured indirectly by PET measurements of \hta
receptor binding and serotonin transporter binding.  In agreement with
animal and experimental studies, we were able to show a non-linear
association between these two serotonergic markers.  Our model
\correction{describes}{represents} a first step towards linking several measurements of the
serotonergic system into a simultaneous description of central 5-HT
levels.  An interesting longer-term perspective of this model is the
\correction{application of it}{possibility} to \correction{describe}{explore} the association between latent 5-HT
levels and the development of neuropsychiatric diseases such as major
depressive episodes. The extension of \twostage
to allow for binary and time-to-event endpoints will be
\correction{the}{a} topic for future research.


\vspace*{1em}
\begin{center}
\emph{Acknowledgements}
\end{center}
The project is part of the Dynamical Systems Interdisciplinary Network,
University of Copenhagen and it has also recived support from
Innovation Fund Denmark.  The authors wish to thank the staff at the
Neurobiology Research Unit, Rigshospitalet, for providing the data
example and help with the interpretation of the data.

\bibliography{nonref}
\bibliographystyle{apalike}

\clearpage
\appendix
\section{General proof of consistency and robustness to direct
  associations between stage one and stage two model}\label{sec:app:consistent}

In this section, we prove that two-stage estimation is consistent also in the
situation where some covariates may have direct effects on the
outcomes of the stage 2 model (\(K \ne 0\)). We also show that the
two-stage estimator can be modified to yield consistent estimation in
situations where there are direct effects of outcomes in the stage 1 model on
outcomes  in the stage 2 model.

\begin{thm}\label{thm:consistencya}
  Under the assumptions of the non-linear structural equation
  model with \(K \ne 0\), the two-stage estimator will yield consistent estimation of all 
  parameters (\(\theta\))  except for the residual covariance, \(\Psi\), of the latent
  variables in step 2.
  \begin{proof}
    Since the exposure part of
    the model is correctly specified \(\theta_1\) is estimated
    using  ML-estimation
    in the marginal distribution of \(X\) given \(Z\) and therefore
    \(\widehat{\theta}_1\) is consistent under mild regularity
    conditions \citep{anderson_normalfactor_annals1988}.\\

    In step 2, the model is misspecified as we fit a linear model to a
    non-linear association. In a linear SEM, the estimator will converge
    to the parameter value inducing the modelled  
    mean and covariance which is closest to the true mean and
    covariance \cite{white1982maximum}. We prove the theorem by showing that even though 
    the model is misspecified it includes
    the true mean and covariance of the data. Therefore, \(\widehat{\theta}_2\) will
    converge to the parameter value which induces the true mean and
    variance. Finally, this value is
    characterized and it is seen to be
    identical to the truth (\(\theta_2\))
    except for elements describing the  residual
    covariance of the latent variables in step 2 (\(\Psi\)). \\

    In step 2, we fit a linear model to  the data 
    \(V_i = (Z_i^t,\predphi(\xi_i)^t, Y_i^t)^t\). 
    It is assumed that this model has the correct measurement part
    (\(Y_i  =  {\nu_2} + \Lambda_2 \eta_i  + K_2
    Z_i+\epsilon_i\)). Here the subscript 2 is used to distinguish the
    parameters of model 2 from the true parameter value (no subscript). 
    The structural model is: \(
    \eta_i  =  \alpha_2 + B_2\predphi(\xi_i) + \Gamma_2 Z_i + \zeta_i\).
    Using standard results from linear  models \cite{bollen1989structural},
    we now derive expressions for the modelled mean and variance

    \small
\begin{eqnarray}\label{ema}
 \begin{array}{ccl}
        \mu_{V_i} & = &[\mu_Z,
        \mu_{\varphi}, \nu_2 + \Lambda_2
                                                   \alpha_2 + \Lambda_2 B_2  \mu_{\varphi} + (\Lambda_2 B_2 +  K_2) \mu_Z ]
                                                     
      \end{array}
  \end{eqnarray}

 \begin{eqnarray}\label{varma}
 \begin{array}{l}
        \Sigma_{V_i}
                                               = \\
                                                 
                                                   \left( \begin{array}{ccc}

                                                              \Sigma_Z
                                                              &
                                                              \Sigma_{z,{\varphi}}
                                                              &
                                                 
                                                              [
                                                              \Sigma_{{\varphi},z} B_2^t
                                                              +
                                                              
                                                              \Sigma_z\Gamma_2^t] \Lambda_2^t
                                                              +
                                                            \Sigma_Z K_2^t\\
                                                             . &
                                                            \Sigma_{\varphi}
                                                            &
                                                            [
                                                            \Sigma_{\varphi} B_2^t
                                                            + 
                                                            \Sigma_{{\varphi},z} \Gamma_2^t
                                                            ] \Lambda_2^t+ 
                                                            \Sigma_{{\varphi},z}
                                                            K_2^t\\
                                                              . &
                                                             . & \Lambda_2
                                                            \Sigma_{\eta}
                                                            \Lambda_2^t
                                                            +
                                                            \Lambda_2
                                                            \Gamma_2\Sigma_zK_2^t+K_2\Sigma_z\Gamma_2^t\Lambda_2^t+
                                                              \Lambda_2
                                                            B_2\Sigma_{{\varphi},z}
                                                            K_2^t
                                                            +
                                                            K_2\Sigma_{z,{\varphi}}
                                                            B_2^t\Lambda_2^t+
                                                        K_2\Sigma_zK_2^t+ \Omega_2 
                                                      \end{array}\right)
      \end{array}
  \end{eqnarray}

\normalsize
    where 
$\Sigma_{\eta}
        =  B_2 \Sigma_{\varphi} B_2^t + B_2
              \Sigma_{{\varphi},z} \Gamma_2^t +
              \Gamma_2\Sigma_{z,{\varphi}} B_2^t + \Gamma_2
              \Sigma_Z \Gamma_2^t + \Psi_2$ 
    is the modelled variance of \(\eta\) and \(\mu_{\varphi}, \mu_{Z}, \Sigma_{\varphi},
    \Sigma_{{\varphi},z}, \Sigma_Z\) are completely unstructured 
    parameters modelling the mean and variance of the step-two covariates
    (\(\predphi(\xi_i), Z\)). \\

    To calculate
    the true mean and variance of   \(V_i\) we re-write the structural
    model as 
    \begin{eqnarray}\label{stmodelaa}
      \begin{array}{lcl} 
        \eta_i & = & \alpha + B \varphi(\xi_i) + \Gamma Z_i + \zeta_i\\
               & = & \alpha + B\predphi(\xi_i)   + B [\varphi(\xi_i)-
                     \predphi(\xi_i)] + \Gamma Z_i + \zeta_i,
      \end{array} 
    \end{eqnarray}

From the law of iterated expectations it follows that 
\(\E[\varphi(\xi_i)-\predphi(\xi_i)]=0\) and therefore
 the
    marginal mean of the data is  
\begin{eqnarray}\label{eta}
      \begin{array}{ccl}
        \E\{V_i\} & = &[\E(Z_i),
        \E\{\varphi(\xi_i)\},
\nu + \Lambda \alpha +
                                                   \Lambda B
                                                   \E\{\varphi(\xi_i)\}+(\Lambda
                                                   \Gamma + K)
                                                   \E(Z_i)]
\end{array}
    \end{eqnarray}

For calculation of the variance,
we first derive variance of the latent variable \(\eta_i\) again using
equation (\ref{stmodelaa})
    \begin{eqnarray}\label{varetata}
      \begin{array}{lcl} 
        \var(\eta_i) &= & B \var [\predphi(\xi_i)] B^t + B 
                          \cov[\predphi(\xi_i), Z_i] \Gamma^t +  \Gamma \cov[Z_i,
                          \predphi(\xi_i)] B^t\\
                     & & + B \var[\varphi(\xi_i)- \predphi(\xi_i)]B^t + \Gamma \var(Z_i) \Gamma^t + 
                         \Psi.
      \end{array} 
    \end{eqnarray}

    because
    \(\cov[\predphi(\xi_i), \varphi(\xi_i)-
    \predphi(\xi_i)]=0\) and  \(\cov[Z_i, \varphi(\xi_i)-
    \predphi(\xi_i)]=0\), which again follows from  
 the law of iterated expectations.\\

Now the
    variance of the step-two data is 
    \begin{eqnarray}\label{varta}
      \begin{array}{l}
        \var[V_i] = \\
                                                 
                                                   \left( \begin{array}{lll}
                                                           \var(Z_i) &
                                                           \cov[Z_i,\predphi(\xi_i)] &
                                                           [\cov\{\predphi(\xi_i),
                                                           Z_i\} B^t+
                                                           \var(Z_i)
                                                           \Gamma^t] \Lambda^t
                                                           + 
                                                           \var(Z_i) K^t\\
                                                            . &
                                                            \var[\predphi(\xi_i)]
                                                            &  
                                                          [\var
                                                            \{\predphi(\xi_i)\}B^t  + 
                                                            \cov\{\predphi(\xi_i),Z_i\}\Gamma^t
                                                            ]
                                                            \Lambda^t
                                                            +
                                                            \cov\{\predphi(\xi_i),Z_i\}
                                                            K^t\\
                                                             . &  . & 
                                                           \var(Y_i) 
                                                          \end{array}\right)
      \end{array}
    \end{eqnarray}\\

    \normalsize
where \(\var(Y_i)=\Lambda \var(\eta_i)
                                                            \Lambda^t
                                                            + \Lambda
                                                            \Gamma\var(Z_i)K^t+K\var(Z_i)\Gamma^t\Lambda^t+ \Lambda
                                                            B \cov[\predphi(\xi_i),
                                                           Z_i]K^t+K \cov[Z_i,\predphi(\xi_i)] B^t\Lambda^t+ 
K\var(Z_i)K^t+
                                                           \Omega\).
 Here we used
    \(\cov[\predphi(\xi_i),Y_i]=\cov[\predphi(\xi_i),\Lambda \eta_i] = 
\cov[\predphi(\xi_i), B
    \predphi(\xi_i)+ \Gamma Z_i] \Lambda^t\) as \(\cov[\predphi(\xi_i), \varphi(\xi_i)-
    \predphi(\xi_i)]=0\)
    and
\(\cov(Z_i,Y_i)=\cov(Z_i\Lambda \eta_i)= \cov[Z_i,B
    \predphi(\xi_i)+ \Gamma Z_i] \Lambda^t\) as \(\cov[Z_i, \varphi(\xi_i)-
    \predphi(\xi_i)]=0\).\\

Now it is straight forward to see that the modelled mean and covariance
    is equal to the corresponding true values [(\ref{ema}) =
    (\ref{eta}) and (\ref{varma}) =
    (\ref{varta})]
if \(B_2=B,
    \Lambda_2=\Lambda, \Gamma_2=\Gamma, \Omega_2=\Omega, \alpha_2= \alpha,
    \nu_2=\nu, \mu_{\varphi}= \E[\varphi(\xi_i)], \mu_Z=\E(Z_i),
    \Sigma_{z,{\varphi}}=\cov[Z_i,\predphi(\xi_i)],
    \Sigma_{\varphi}= \var[\predphi(\xi_i)], 
    \Sigma_Z=\var(Z_i)\) 
    and  \(\Psi_2=\Psi+B \var[\varphi(\xi_i)-
    \predphi(\xi_i)] B^t\). Note that, although the expression for
    the modelled variance of the latent variable (\(\Sigma_{\eta}\))
    is missing the term 
\(B \var[\varphi(\xi_i)- \predphi(\xi_i)] B^t\), 
the model  can achieve the correct
    variance of \(Y_i\) 
    by adding the missing term to the residual variance \(\Psi_2\). 
    Since the model
    includes the true mean and variance, \(\widehat{\theta}_2\) will
    converge to the specific parameter value 
inducing the true mean and variance. This means that all parameters are
    consistently estimated 
    except for the variance of the latent variable,
    which will be overestimated. \\

    The above arguments were made for predicted values \(\predphi\).
    The results now follows from the Continuous Mapping Theorem and by
    noting that \(\predphi_n\longrightarrow\predphi \ a.s.\) 
    as \(n\to\infty\).
  \end{proof}
\end{thm}

\begin{figure}[htbp!]
  \centering
  \includegraphics[]{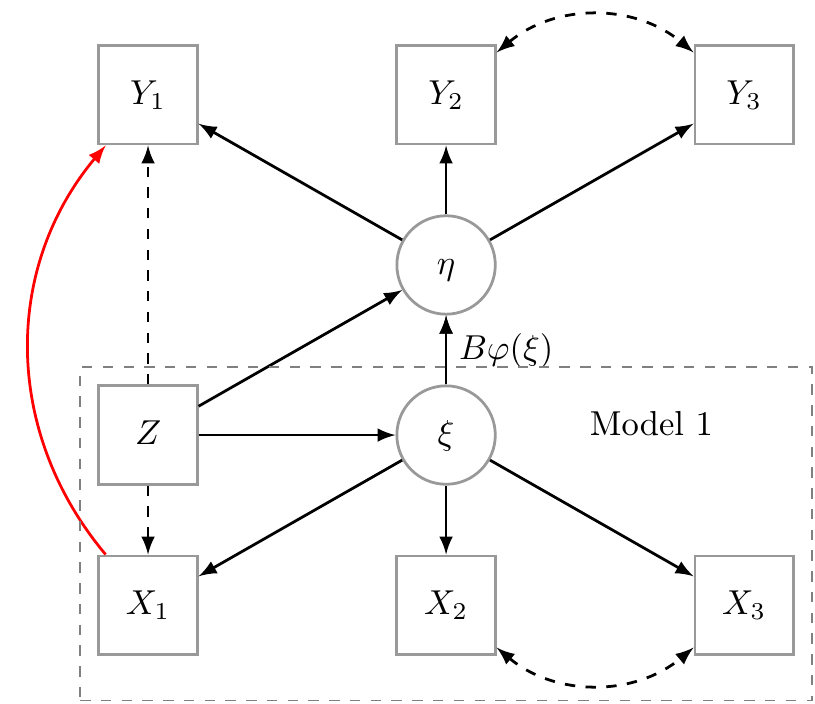}
  \caption{Path diagram showing an example of the non-linear
structural equation
    model with a direct effect of an indicator variable of the stage
    one model onto a variable in the stage two model.
    \label{fig:path1b}}
\end{figure}

\begin{cor}
  Consider a model where there may be direct effects of response
  variables in the stage 1 model on response variables in the stage 2
  model (see Figure \ref{fig:path1b}), i.e., the measurement model of
  stage-two data is
  \(Y_i = \nu + \Lambda \eta_i + K Z_i+ L X_i + \epsilon_i\). Consistent estimation can
  be obtained with the two-stage estimator if the measurement model
  for \(Y_i\) is correctly specified, i.e., \(X_i\) is included in the
  stage 2 analysis as additional covariates.
  \begin{proof}
    The stage-one model is correctly specified so this stage remains
    consistent. According to Theorem \ref{thm:consistencya}, the second
    stage analysis is also consistent when some covariates have direct
    effects on the response variables. The only requirement is that
    these covariates are also included in the first stage analysis, so
    that \(\cov[\varphi(\xi_i)-\predphi(\xi_i), Z_i] =0\). Response
    variables in stage one are of course a part of the stage one
    analysis. A consequence of this is that
    \(\cov[\varphi(\xi_i)-\predphi(\xi_i), X_i] =0\) and therefore
    the stage two analysis is consistent.
  \end{proof}
\end{cor}

\section{Bollen's IV method}
\label{sec:bolleniv}
The idea is to replace latent variables with observed reference
variables and then estimate parameters using regression
techniques. For illustration, we use our simulation model with a
linear structural equation \(\eta=\alpha + \beta \xi + \zeta\). If we use
\(Y_1\) and \(X_1\) as reference indicators, we can replace the latent
variables with observed reference indicators yielding
\(Y_1=\alpha + \beta X_1 -\beta \tilde{\epsilon}_1 +\epsilon_1+ \zeta\).  This creates an
observed-variable-equation, where predictors are correlated with error
terms and therefore least squares estimation would be biased, but a
consistent analysis can be performed by using non-reference indicators
as instrumental variables. In our example the indicators \(X_2\) and
\(X_3\) could be used as instruments for \(X_1\). In linear models,
Bollen have shown that this IV-estimator is more robust to model
misspecification than ML-analysis \citep{bollen2007latent}. He has also
extended the method to non-linear models (\cite{bollen1995structural,
  bollen1998interactions}). Here instrumental variables
must be identified for the non-linear terms and this may be
challenging. For a quadratic model (\(\eta=\alpha + \beta \varphi(\xi)
+ \zeta\) with \(\varphi(\xi)=\xi^2\)), Bollen showed that consistent estimation can be achieved by
replacing $\xi$ by $X_1$ and by using \emph{transformed} variables
\(X_2^2 \) and \(X_3^2\)  as instruments for the non-linear effect term. 
It is not clear that this procedure would generally yield
consistent estimation, but for product and quadratic terms, where
\(\varphi(X_1-\tilde{\epsilon}_1)=\varphi(X_1)+\varphi_1(X_1, \tilde{\epsilon}_1)
\), the observed-variable-equation takes the form
\(Y_1=\alpha + \beta \varphi(X_1) + error\), and IV estimation will be
consistent as long as it is possible to find instrumental variables
that are uncorrelated with the error term.  
For standard errors, Bollen proposed the usual
IV-solution, which in our example takes the form
\(\widehat{\sigma}_e^2 (\widehat{X}^t \widehat{X} )^{-1}\), where
\(\widehat{X}\) is $n\times 2$-matrix with  an
intercept column of ones and a column
of predicted values of \(X_1\) obtained from a regression on
\(X_2,X_3\), while \(\widehat{\sigma}_e^2= \sum_{i=1}^n u_i^2/n\) with residuals
\(u_i=Y_{i1}-\widehat{\alpha} - \widehat{\beta}X_{i1}\). However, this
expression assumes homoskedastic errors, which is generally not
satisfied in non-linear SEMs. For example in the quadratic model,
the observed-variable-equation becomes
\(Y_1=\alpha + \beta X_1^2 -\beta \tilde{\epsilon}_1 X_1 + \beta \tilde{\epsilon}_1^2 +\epsilon_1+ \zeta\) and the error
term
(\(\beta \tilde{\epsilon}_1 X_1 + \beta \tilde{\epsilon}_1^2
+\epsilon_1+ \zeta\)) will have a variance that
depends on \(X_1\).
Instead, we suggest using the heteroskedasticity robust standard errors
for the IV \citep{wooldridge2010econometric}, i.e.
\((\widehat{X}^t \widehat{X} )^{-1} \widehat{X}^t S \widehat{X}
(\widehat{X}^t \widehat{X} )^{-1}\), where \(S\) is a diagonal matrix
of squared residuals (\(u_i^2\)).

\clearpage
\definecolor{mygray}{rgb}{0.2,0.2,0.2}
\DefineVerbatimEnvironment{verbatim}{Verbatim}{fontsize=\footnotesize, xleftmargin=2em, formatcom=\color[rgb]{0.3,0.3,0.45}}
\lstset{
  breakatwhitespace=false,         
  breaklines=true,                 
  deletekeywords={*.~,<,>,-,...},            
  escapeinside={(@}{@)},          
  extendedchars=true,              
  keepspaces=true,                 
  stepnumber=1,                    
  numbers=left,                    
  numbersep=6pt,                   
  numberstyle=\tiny\color{gray}, 
  frame=lines,	                   
  tabsize=2,	                   
  title=\lstname                   
  rulecolor=\color{black},         
  showspaces=false,                
  showstringspaces=false,          
  showtabs=false,                  
  captionpos=b,                    
  basicstyle=\color{mygray}\linespread{0.95}\small\ttfamily,
  ndkeywordstyle=\color{gray},
  keywordstyle=\color{black}\bfseries,
  commentstyle=\color{gray}\ttfamily\itshape,
  stringstyle=\color{gray}\itshape,
  alsoletter=.,
  numbers=left,
  xleftmargin=.23in,
  aboveskip=1.5em,
  belowskip=1em,
  columns=fullflexible,
  backgroundcolor=\color{white},
  basewidth={0.5em,0.42em},
  language=r,
  morendkeywords={NA,TRUE,FALSE,Inf}
  morekeywords={lvm,functional,lvm,sim,nonlinear,...}            
}
\lstdefinestyle{rplus}{language=r,
otherkeywords={<-,\$}
morendkeywords={NA,TRUE,FALSE,Inf}
morekeywords={lvm,functional,lvm,sim,nonlinear,...}            
}

\section{Software implementation}
\label{appendix:software}

The two-stage estimator is implemented in the \texttt{lava} package in the
statistical software R citep:Rlang. The source code is hosted
at GitHub and can be downloaded freely from
\texttt{https://github.com/kkholst/lava}. The version on which this paper is
based on (doi: \texttt{10.5281/zenodo.1411865}) can be installed with the following command (depending on the
\texttt{devtools} package available from CRAN):

\lstset{numbers=left,language=r,label= ,caption= ,captionpos=b,morekeywords={install_github},alsoletter={_},deletekeywords={*},otherkeywords={<-,\$}}
\begin{lstlisting}
devtools::install_github("kkholst/lava", ref="twostage")
\end{lstlisting}

To demonstrate the syntax we simulate from the following measurement models (see Figure ref:fig:lvm1) 
\begin{figure}[htbp]
\centering
\includegraphics[width=0.6\textwidth]{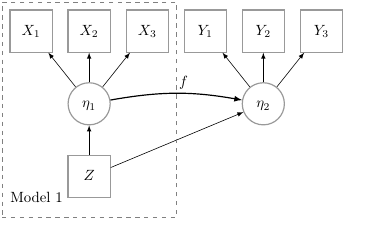}
\caption{Path diagram of the simulation model specified by (\ref{ex:measurements})-(\ref{ex:eta1}). label:fig:lvm1}
\end{figure}

\begin{align}\label{ex:measurements}
\begin{split}
X_{j} &= \eta_{1} + \epsilon_{j}^{x}, \quad j=1,2,3 \\
Y_{j} &= \eta_{2} + \epsilon_{j}^{y}, \quad j=1,2,3
\end{split}
\end{align}
and with a structural model given by
\begin{align}
\eta_{2} &= f(\eta_{1}) + Z + \zeta_{2}\label{ex:eta2} \\
\eta_{1} &= Z + \zeta_{1}\label{ex:eta1}
\end{align}
with iid measurement errors
\(\epsilon_{j}^{x},\epsilon_{j}^{y},\zeta_{1},\zeta_{2}\sim\mathcal{N}(0,1),
j=1,2,3.\) and standard normal distributed covariate \(Z\).  To
simulate from this model we use the following syntax:

\lstset{numbers=left,language=r,label= ,caption= ,captionpos=b,morekeywords={library},otherkeywords={<-,\$}}
\begin{lstlisting}
library("lava")
\end{lstlisting}

\begin{verbatim}
lava version 1.6.3
\end{verbatim}

\lstset{numbers=left,language=r,label= ,caption= ,captionpos=b,morekeywords={regression,lvm,functional,sim},deletekeywords={*},otherkeywords={<-,\$}}
\begin{lstlisting}
f <- function(x) cos(1.25*x) + x - 0.25*x^2
m <- lvm(x1+x2+x3 ~ eta1, y1+y2+y3 ~ eta2, latent=~eta1+eta2)
regression(m) <- eta1+eta2 ~ z
functional(m, eta2~eta1) <- f

d <- sim(m, n=200, seed=42) # Default is all parameters are 1
\end{lstlisting}

We refer to cite:holstjoergensen\(_{\text{lava}}\) for details on the syntax for
model specification.
Given the data the first step is now to specify the measurement models in (ref:ex:measurements):
\lstset{numbers=left,language=r,label= ,caption= ,captionpos=b,morekeywords={lvm},otherkeywords={<-,\$}}
\begin{lstlisting}
m1 <- lvm(x1+x2+x3 ~ eta1, eta1 ~ z, latent=~eta1)
m2 <- lvm(y1+y2+y3 ~ eta2, eta2 ~ z, latent=~eta2)
\end{lstlisting}

Next, we specify a quadratic relationship between the two latent variables
\lstset{numbers=left,language=r,label= ,caption= ,captionpos=b,morekeywords={nonlinear},otherkeywords={<-,\$}}
\begin{lstlisting}
nonlinear(m2, type="quadratic") <- eta2 ~ eta1
\end{lstlisting}

and the model can then be estimated using the two-stage estimator
\lstset{numbers=left,language=r,label= ,caption= ,captionpos=b,morekeywords={twostage},otherkeywords={<-,\$}}
\begin{lstlisting}
e1 <- twostage(m1, m2, data=d)
e1
\end{lstlisting}

\begin{verbatim}
                    Estimate Std. Error  Z-value   P-value
Measurements:                                             
   y2~eta2           0.97686    0.03451 28.30865    <1e-12
   y3~eta2           1.04485    0.03485 29.98153    <1e-12
Regressions:                                              
   eta2~z            0.88513    0.20778  4.25997 2.045e-05
   eta2~eta1_1       1.14072    0.17410  6.55194 5.679e-11
   eta2~eta1_2      -0.45055    0.07161 -6.29199 3.134e-10
Intercepts:                                               
   y2               -0.12198    0.10915 -1.11749    0.2638
   y3               -0.09879    0.10545 -0.93680    0.3489
   eta2              0.67814    0.17363  3.90567 9.397e-05
Residual Variances:                                       
   y1                1.30730    0.17743  7.36790          
   y2                1.11056    0.14478  7.67064          
   y3                0.80961    0.13203  6.13219          
   eta2              2.08483    0.28985  7.19274
\end{verbatim}

We see a clear statistically significant effect of the second order
term (\texttt{eta2\textasciitilde{}eta1\_2}). For comparison, we can also estimate the full MLE
of the linear model:
\lstset{numbers=left,language=r,label= ,caption= ,captionpos=b,morekeywords={estimate,regression},otherkeywords={<-,\$}}
\begin{lstlisting}
e0 <- estimate(regression(m1%++%m2, eta2~eta1), d)
estimate(e0,keep="^eta2~[a-z]",regex=TRUE) ## Extract coef. matching reg.ex.
\end{lstlisting}

\begin{verbatim}
          Estimate Std.Err    2.5% 97.5%   P-value
eta2~eta1   1.4140  0.2261 0.97083 1.857 4.014e-10
eta2~z      0.6374  0.2778 0.09291 1.182 2.177e-02
\end{verbatim}

Next, we calculate predictions from the quadratic model using the estimated parameter coefficients
\[
\E_{\widehat{\theta}_{2}}(\eta_{2} \mid \eta_{1}, Z=0),
\]
\lstset{numbers=left,language=r,label= ,caption= ,captionpos=b,morekeywords={Col,head},deletekeywords={data,by,col},otherkeywords={<-,\$}}
\begin{lstlisting}
newd <- expand.grid(eta1=seq(-4, 4, by=0.1), z=0)
pred1 <- predict(e1, newdata=newd, x=TRUE)
head(pred1)
\end{lstlisting}

\begin{verbatim}
             y1         y2         y3       eta2
[1,] -11.093569 -10.958869 -11.689950 -11.093569
[2,] -10.623561 -10.499736 -11.198861 -10.623561
[3,] -10.162565 -10.049406 -10.717187 -10.162565
[4,]  -9.710579  -9.607878 -10.244928  -9.710579
[5,]  -9.267605  -9.175153  -9.782084  -9.267605
[6,]  -8.833641  -8.751230  -9.328656  -8.833641
\end{verbatim}

To obtain a potential better fit we next proceed with a natural cubic spline 
\lstset{numbers=left,language=r,label= ,caption= ,captionpos=b,morekeywords={twostage,nonlinear},deletekeywords={c},otherkeywords={<-,\$}}
\begin{lstlisting}
kn <- seq(-3,3,length.out=5)
nonlinear(m2, type="spline", knots=kn) <- eta2 ~ eta1
e2 <- twostage(m1, m2, data=d)
e2
\end{lstlisting}

\begin{verbatim}
                    Estimate Std. Error  Z-value   P-value
Measurements:                                             
   y2~eta2           0.97752    0.03455 28.29248    <1e-12
   y3~eta2           1.04508    0.03488 29.96248    <1e-12
Regressions:                                              
   eta2~z            0.86729    0.20273  4.27795 1.886e-05
   eta2~eta1_1       2.86231    0.67270  4.25495 2.091e-05
   eta2~eta1_2       0.00344    0.10097  0.03409    0.9728
   eta2~eta1_3      -0.26270    0.29398 -0.89360    0.3715
   eta2~eta1_4       0.50778    0.35191  1.44293     0.149
Intercepts:                                               
   y2               -0.12185    0.10922 -1.11563    0.2646
   y3               -0.09874    0.10545 -0.93638    0.3491
   eta2              1.83814    1.66416  1.10454    0.2694
Residual Variances:                                       
   y1                1.31286    0.17750  7.39647          
   y2                1.10412    0.14455  7.63850          
   y3                0.81124    0.13185  6.15286          
   eta2              1.99404    0.27004  7.38416
\end{verbatim}

Confidence limits can be obtained via the Delta method using the \texttt{estimate} method:
\lstset{numbers=left,language=r,label= ,caption= ,captionpos=b,morekeywords={predict,cbind,estimate,head},deletekeywords={data,by,col},otherkeywords={<-,\$}}
\begin{lstlisting}
p <- cbind(eta1=newd$eta1,
	  estimate(e2,f=function(p) predict(e2,p=p,newdata=newd))$coefmat)
head(p)
\end{lstlisting}

\begin{verbatim}
   eta1  Estimate   Std.Err      2.5%     97.5%      P-value
p1 -4.0 -9.611119 1.2650975 -12.09066 -7.131573 3.027543e-14
p2 -3.9 -9.324887 1.2054915 -11.68761 -6.962167 1.031268e-14
p3 -3.8 -9.038656 1.1467339 -11.28621 -6.791099 3.219580e-15
p4 -3.7 -8.752425 1.0889618 -10.88675 -6.618099 9.176275e-16
p5 -3.6 -8.466193 1.0323409 -10.48954 -6.442842 2.384613e-16
p6 -3.5 -8.179962 0.9770711 -10.09499 -6.264938 5.668675e-17
\end{verbatim}

The fitted function can be obtained with the following code (see Figure ref:fig:pred2b):
\lstset{numbers=left,language=r,label=fig:pred2,caption= ,captionpos=b}
\begin{lstlisting}
plot(I(eta2-z) ~ eta1, data=d, col=Col("black",0.5), pch=16,
     xlab=expression(eta[1]), ylab=expression(eta[2]), xlim=c(-4,4))
lines(Estimate ~ eta1, data=as.data.frame(p), col="darkblue", lwd=5)
confband(p[,1], lower=p[,4], upper=p[,5], polygon=TRUE, 
	 border=NA, col=Col("darkblue",0.2))
\end{lstlisting}

\begin{figure*}[!htb]
\centering
\includegraphics[width=0.9\textwidth]{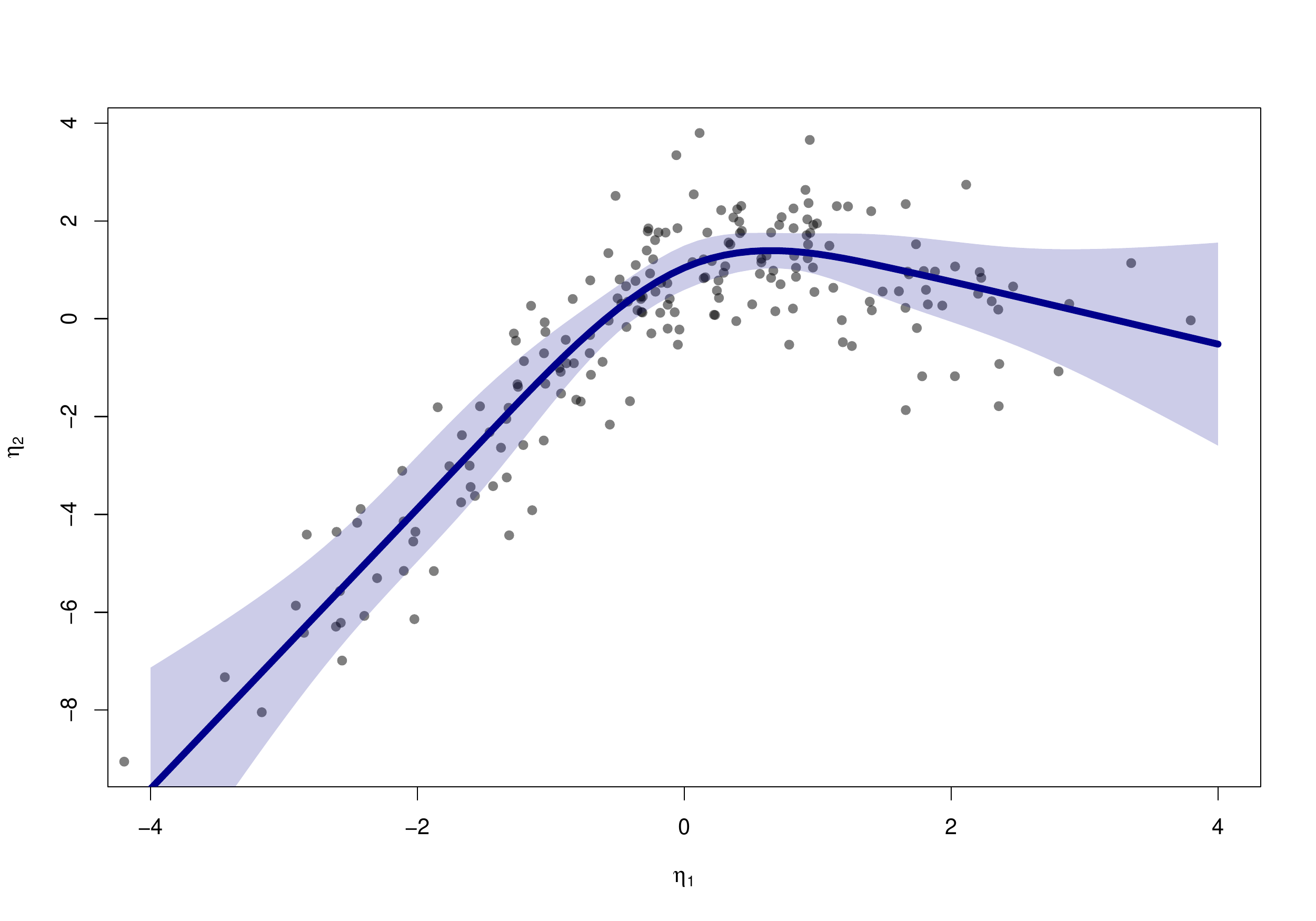}
\caption{Estimated association between \(\eta_{1}\) and \(\eta_{2}\) using natural cubic spline (4 knot points) with point-wise 95\% confidence limits. label:fig:pred2b}
\end{figure*}

\subsection{Cross-validation}

A more formal comparison of the different models can be obtained by
cross-validation. Here we specify linear, quadratic and cubic
spline models with 4 and 9 degrees of freedom.
\lstset{numbers=left,language=r,label= ,caption= ,captionpos=b,morekeywords={nonlinear,endogenous,cv,twostage},deletekeywords={length},otherkeywords={<-,\$}}
\begin{lstlisting}
m2a <- nonlinear(m2, type="linear", eta2~eta1)
m2b <- nonlinear(m2, type="quadratic", eta2~eta1)
kn1 <- seq(-3,3,length.out=5)
kn2 <- seq(-3,3,length.out=8)
m2c <- nonlinear(m2, type="spline", knots=kn1, eta2~eta1)
m2d <- nonlinear(m2, type="spline", knots=kn2, eta2~eta1)
\end{lstlisting}

To assess the model fit average RMSE is estimated with 5-fold cross-validation repeated two times
\lstset{numbers=left,language=r,label= ,caption= ,captionpos=b,morekeywords={nonlinear,endogenous,cv,twostage},deletekeywords={control,start,rep,scale,data,by,col,polygon,lower,upper,expression},otherkeywords={<-,\$}}
\begin{lstlisting}
## Scale models in stage 2 to allow for a fair RMSE comparison
d0 <- d
for (i in endogenous(m2)) 
    d0[,i] <- scale(d0[,i],center=TRUE,scale=TRUE)
## Repeated 5-fold cross-validation:
ff <- lapply(list(linear=m2a,quadratic=m2b,spline4=m2c,spline6=m2d),
	    function(m) function(data,...) twostage(m1,m,data=data,stderr=FALSE,control=list(start=coef(e0),contrain=TRUE)))
fit.cv <- cv(ff,data=d,K=5,rep=2,mc.cores=4,seed=1)
fit.cv
\end{lstlisting}

\begin{verbatim}
              RMSE
linear    4.508896
quadratic 3.270098
spline4   3.105159
spline6   3.376329
\end{verbatim}

Here the RMSE is in favour of the splines model with 4 degrees of freedom (see Figure ref:fig:multifit).

\lstset{numbers=left,language=r,label=multifit,caption= ,captionpos=b}
\begin{lstlisting}
fit <- lapply(list(m2a,m2b,m2c,m2d),
	     function(x) {
		 e <- twostage(m1,x,data=d)
		 pr <- cbind(eta1=newd$eta1,predict(e,newdata=newd$eta1,x=TRUE))
		 return(list(estimate=e,predict=as.data.frame(pr)))
	     })

plot(I(eta2-z) ~ eta1, data=d, col=Col("black",0.5), pch=16,
     xlab=expression(eta[1]), ylab=expression(eta[2]), xlim=c(-4,4))
col <- c("orange","darkred","darkgreen","darkblue")
lty <- c(3,4,1,5)
for (i in seq_along(fit)) {
    with(fit[[i]]$pr, lines(eta2 ~ eta1, col=col[i], lwd=4, lty=lty[i]))
}
legend("bottomright",
      c("linear","quadratic","spline(df=4)","spline(df=6)"),
      col=col, lty=lty, lwd=3)
\end{lstlisting}

\begin{figure*}[!htb]
\centering
\includegraphics[width=0.9\textwidth]{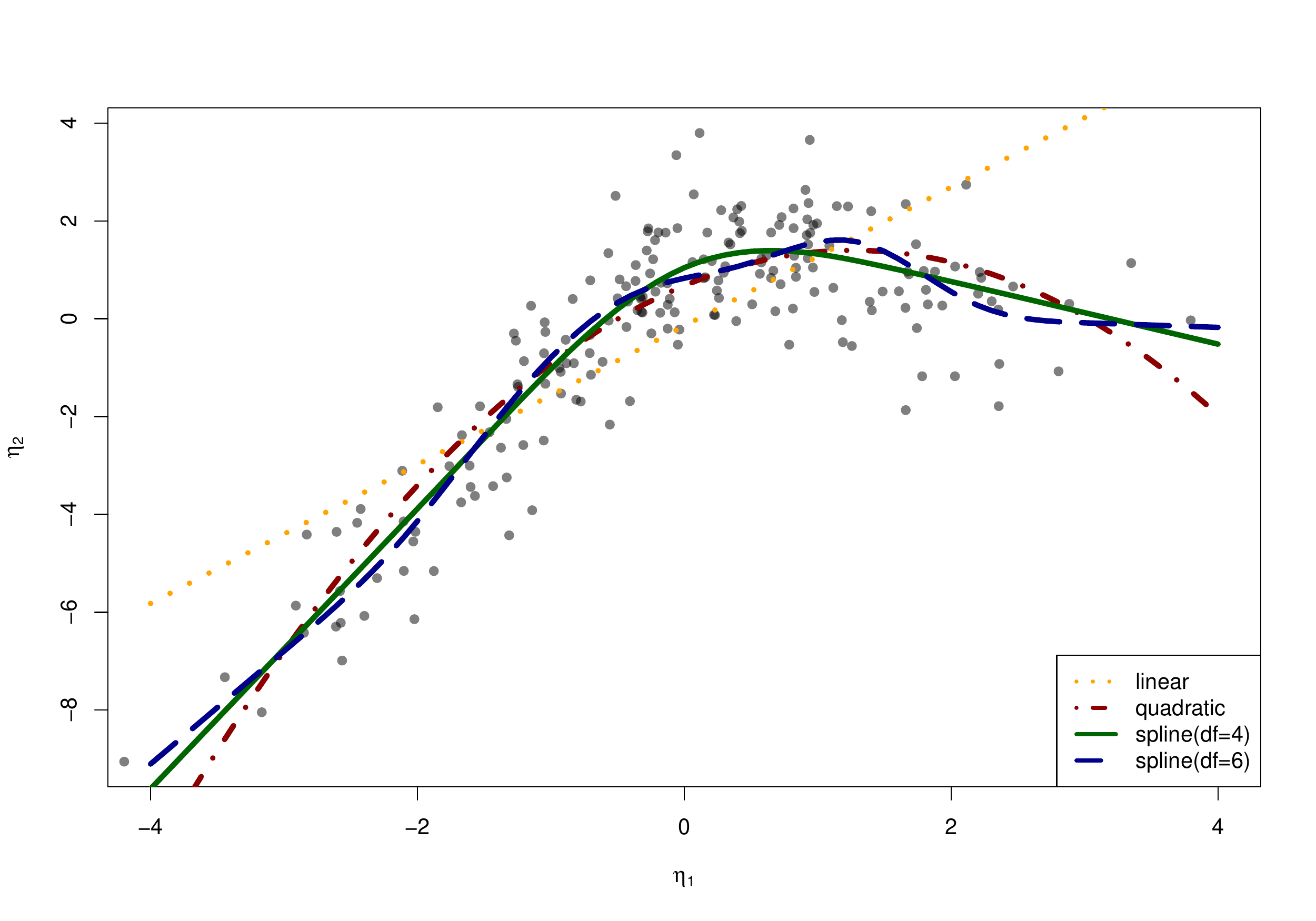}
\caption{Comparison of model fit using linear, quadratic and natural splines models with 4 and 6 degrees of freedom. The points show the actual latent variables. label:fig:multifit}
\end{figure*}

For convenience, the function \texttt{twostageCV} can be used to do the
cross-validation. For example,
\lstset{numbers=left,language=r,label= ,caption= ,captionpos=b}
\begin{lstlisting}
selmod <- twostageCV(m1, m2, data=d, df=2:6, nmix=1:3, 
	  nfolds=5, rep=1, mc.cores=parallel::detectCores())
\end{lstlisting}
applies 5-fold cross-validation to select the best splines with
degrees of freedom varying from 1-6 (the linear model is
automatically included)

\lstset{numbers=left,language=r,label= ,caption= ,captionpos=b}
\begin{lstlisting}
selmod
\end{lstlisting}

\begin{verbatim}
______________________________________________________________________
Selected mixture model: 2 components
      AIC1
1 1961.839
2 1958.803
3 1962.046
______________________________________________________________________
Selected spline model degrees of freedom: 2
Knots: -3.958 0.02149 4.001 

     RMSE(nfolds=5, rep=1)
df:1              4.535541
df:2              3.611836
df:3              3.777584
df:4              3.657840
df:5              4.144918
df:6              3.707977
______________________________________________________________________

                    Estimate Std. Error Z-value  P-value 
Measurements:                                            
   y1~eta2           1.00000                             
   y2~eta2           0.97794  0.03463   28.24076   <1e-12
   y3~eta2           1.04520  0.03473   30.09595   <1e-12
Regressions:                                             
   eta2~z            1.02819  0.22297    4.61136 4e-06   
   eta2~eta1_1       3.41773  0.36899    9.26228   <1e-12
   eta2~eta1_2      -0.05122  0.00706   -7.25313   <1e-12
Intercepts:                                              
   y1                0.00000                             
   y2               -0.12176  0.10921   -1.11495 0.2649  
   y3               -0.09872  0.10547   -0.93600 0.3493  
   eta2              3.93712  0.54020    7.28824   <1e-12
Residual Variances:                                      
   y1                1.31625  0.17654    7.45594         
   y2                1.09975  0.14507    7.58079         
   y3                0.81270  0.13258    6.12986         
   eta2              2.01822  0.28971    6.96633
\end{verbatim}

\subsection{Specification of general functional forms}

Next, we show how to specify a general functional relation of
multiple different latent or exogenous variables. This is achieved via
the \texttt{predict.fun} argument. To illustrate this we include interactions
between the latent variable \(\eta_{1}\) and a dichotomized version of
the covariate \(z\)
\lstset{numbers=left,language=r,label= ,caption= ,captionpos=b,morekeywords={nonlinear,regression,lvm,sim,functional,estimate,twostage,Col,confband,expand.grid},deletekeywords={c,\*,\~,*,data,by,col,polygon,lower,upper,var,expression},alsoletter={.,\*,\~,~},otherkeywords={<-,\$}}
\begin{lstlisting}
d$g <- (d$z<0)*1 ## Group variable
mm1 <- regression(m1, ~g)  # Add grouping variable as exogenous variable (effect specified via 'predict.fun')
mm2 <- regression(m2, eta2~ u1+u2+u1:g+u2:g+z)
pred <- function(mu,var,data,...) {
    cbind("u1"=mu[,1],"u2"=mu[,1]^2+var[1],
	  "u1:g"=mu[,1]*data[,"g"],"u2:g"=(mu[,1]^2+var[1])*data[,"g"])
}
ee1 <- twostage(mm1, model2=mm2, data=d, predict.fun=pred)
estimate(ee1,keep="eta2~u",regex=TRUE)
\end{lstlisting}

\begin{verbatim}
          Estimate Std.Err    2.5%   97.5%  P-value
eta2~u1     0.9891  0.3020  0.3971  1.5810 0.001057
eta2~u2    -0.3962  0.1443 -0.6791 -0.1133 0.006047
eta2~u1:g   0.4487  0.4620 -0.4568  1.3543 0.331409
eta2~u2:g   0.0441  0.2166 -0.3804  0.4686 0.838667
\end{verbatim}

A formal test show no statistically significant effect of this interaction
\lstset{numbers=left,language=r,label= ,caption= ,captionpos=b}
\begin{lstlisting}
summary(estimate(ee1,keep="(:g)", regex=TRUE))
\end{lstlisting}

\begin{verbatim}
Call: estimate.default(x = ee1, keep = "(:g)", regex = TRUE)
__________________________________________________
          Estimate Std.Err    2.5%  97.5% P-value
eta2~u1:g   0.4487  0.4620 -0.4568 1.3543  0.3314
eta2~u2:g   0.0441  0.2166 -0.3804 0.4686  0.8387

 Null Hypothesis: 
  [eta2~u1:g] = 0
  [eta2~u2:g] = 0 
 
chisq = 0.9441, df = 2, p-value = 0.6237
\end{verbatim}

\subsection{Mixture models}
\label{sec:mixmod}
Lastly, we demonstrate how the distributional assumptions of stage 1
model can be relaxed by letting the conditional distribution of the
latent variable given covariates follow a Gaussian mixture
distribution. The following code explicitly defines the parameter
constraints of the model by setting the intercept of the first
indicator variable, \(x_{1}\), to zero and the factor loading
parameter of the same variable to one.
\lstset{numbers=left,language=r,label= ,caption= ,captionpos=b}
\begin{lstlisting}
m1 <- baptize(m1)  ## Label all parameters
intercept(m1, ~x1+eta1) <- list(0,NA) ## Set intercept of (@$x_{1}$@) to zero. Remove the label of (@$\eta_{1}$@)
regression(m1,x1~eta1) <- 1 ## Factor loading fixed to 1
\end{lstlisting}

The mixture model may then be estimated using the \texttt{mixture} method,
where the Parameter names shared across the different mixture
components given in the \texttt{list} will be constrained to be identical in
the mixture model. Thus, only the intercept of \(\eta_{1}\) is
allowed to vary between the mixtures.
\lstset{numbers=left,language=r,label= ,caption= ,captionpos=b}
\begin{lstlisting}
em0 <- mixture(list(m1,m1), data=d)
\end{lstlisting}

To decrease the risk of using a local maximizer of the likelihood we
can rerun the estimation with different random starting values
\lstset{numbers=left,language=r,label= ,caption= ,captionpos=b}
\begin{lstlisting}
em0 <- NULL
ll <- c()
for (i in 1:5) {
    set.seed(i)
    em <- mixture(list(m1,m1), data=d, control=list(trace=0))
    ll <- c(ll,logLik(em))
    if (is.null(em0) || logLik(em0)<tail(ll,1))
	em0 <- em
}
\end{lstlisting}

\lstset{numbers=left,language=r,label= ,caption= ,captionpos=b}
\begin{lstlisting}
summary(em0)
\end{lstlisting}

\begin{verbatim}
Cluster 1 (n=162, Prior=0.776):
--------------------------------------------------
                    Estimate Std. Error Z value  Pr(>|z|)
Measurements:                                            
   x1~eta1           1.00000                             
   x2~eta1           0.99581  0.07940   12.54101   <1e-12
   x3~eta1           1.06344  0.08436   12.60542   <1e-12
Regressions:                                             
   eta1~z            1.06675  0.08527   12.50995   <1e-12
Intercepts:                                              
   x1                0.00000                             
   x2                0.03845  0.09890    0.38883 0.6974  
   x3               -0.02549  0.10333   -0.24666 0.8052  
   eta1              0.20923  0.13162    1.58963 0.1119  
Residual Variances:                                      
   x1                0.98539  0.13316    7.40024         
   x2                0.97181  0.13156    7.38698         
   x3                1.01316  0.14294    7.08812         
   eta1              0.29047  0.11129    2.61004         

Cluster 2 (n=38, Prior=0.224):
--------------------------------------------------
                    Estimate Std. Error Z value  Pr(>|z|)
Measurements:                                            
   x1~eta1           1.00000                             
   x2~eta1           0.99581  0.07940   12.54101   <1e-12
   x3~eta1           1.06344  0.08436   12.60542   <1e-12
Regressions:                                             
   eta1~z            1.06675  0.08527   12.50995   <1e-12
Intercepts:                                              
   x1                0.00000                             
   x2                0.03845  0.09890    0.38883 0.6974  
   x3               -0.02549  0.10333   -0.24666 0.8052  
   eta1             -1.44298  0.25868   -5.57821 2.43e-08
Residual Variances:                                      
   x1                0.98539  0.13316    7.40024         
   x2                0.97181  0.13156    7.38698         
   x3                1.01316  0.14294    7.08812         
   eta1              0.29047  0.11129    2.61004         
--------------------------------------------------
AIC= 1958.803 
||score||^2= 2.124942e-06
\end{verbatim}

Measured by AIC there is a slight improvement in the model fit using the mixture model
\lstset{numbers=left,language=r,label= ,caption= ,captionpos=b}
\begin{lstlisting}
e0 <- estimate(m1,data=d)
AIC(e0,em0)
\end{lstlisting}

\begin{verbatim}
    df      AIC
e0  10 1961.839
em0 12 1958.803
\end{verbatim}

The spline model may then be estimated as before with the \texttt{two-stage} method
\lstset{numbers=left,language=r,label= ,caption= ,captionpos=b}
\begin{lstlisting}
em2 <- twostage(em0,m2,data=d)
em2
\end{lstlisting}

\begin{verbatim}
                    Estimate Std. Error  Z-value   P-value
Measurements:                                             
   y2~eta2           0.97823    0.03469 28.19903    <1e-12
   y3~eta2           1.04530    0.03484 30.00720    <1e-12
Regressions:                                              
   eta2~z            1.02886    0.22330  4.60763 4.073e-06
   eta2~eta1_1       2.80407    0.65493  4.28149 1.856e-05
   eta2~eta1_2      -0.02249    0.09996 -0.22495     0.822
   eta2~eta1_3      -0.17333    0.28933 -0.59909    0.5491
   eta2~eta1_4       0.38673    0.33983  1.13801    0.2551
Intercepts:                                               
   y2               -0.12171    0.10925 -1.11407    0.2653
   y3               -0.09870    0.10546 -0.93592    0.3493
   eta2              2.12363    1.66552  1.27505    0.2023
Residual Variances:                                       
   y1                1.31872    0.17657  7.46862          
   y2                1.09691    0.14503  7.56340          
   y3                0.81345    0.13259  6.13507          
   eta2              1.99591    0.28454  7.01450
\end{verbatim}

In practice the results are very similar to the Gaussian model as shown in Figure ref:fig:mixturefit.
\lstset{numbers=left,language=r,label=mixturefit,caption= ,captionpos=b}
\begin{lstlisting}
plot(I(eta2-z) ~ eta1, data=d, col=Col("black",0.5), pch=16,
     xlab=expression(eta[1]), ylab=expression(eta[2]))

lines(Estimate ~ eta1, data=as.data.frame(p), col="darkblue", lwd=5)
confband(p[,1], lower=p[,4], upper=p[,5], polygon=TRUE, 
	 border=NA, col=Col("darkblue",0.2))

pm <- cbind(eta1=newd$eta1,
	    estimate(em2, f=function(p) predict(e2,p=p,newdata=newd))$coefmat)
lines(Estimate ~ eta1, data=as.data.frame(pm), col="darkred", lwd=5)
confband(pm[,1], lower=pm[,4], upper=pm[,5], polygon=TRUE, 
	 border=NA, col=Col("darkred",0.2))
legend("bottomright", c("Gaussian","Mixture"), 
       col=c("darkblue","darkred"), lwd=2, bty="n")  
\end{lstlisting}

\begin{figure*}[!htb]
\centering
\includegraphics[width=0.9\textwidth]{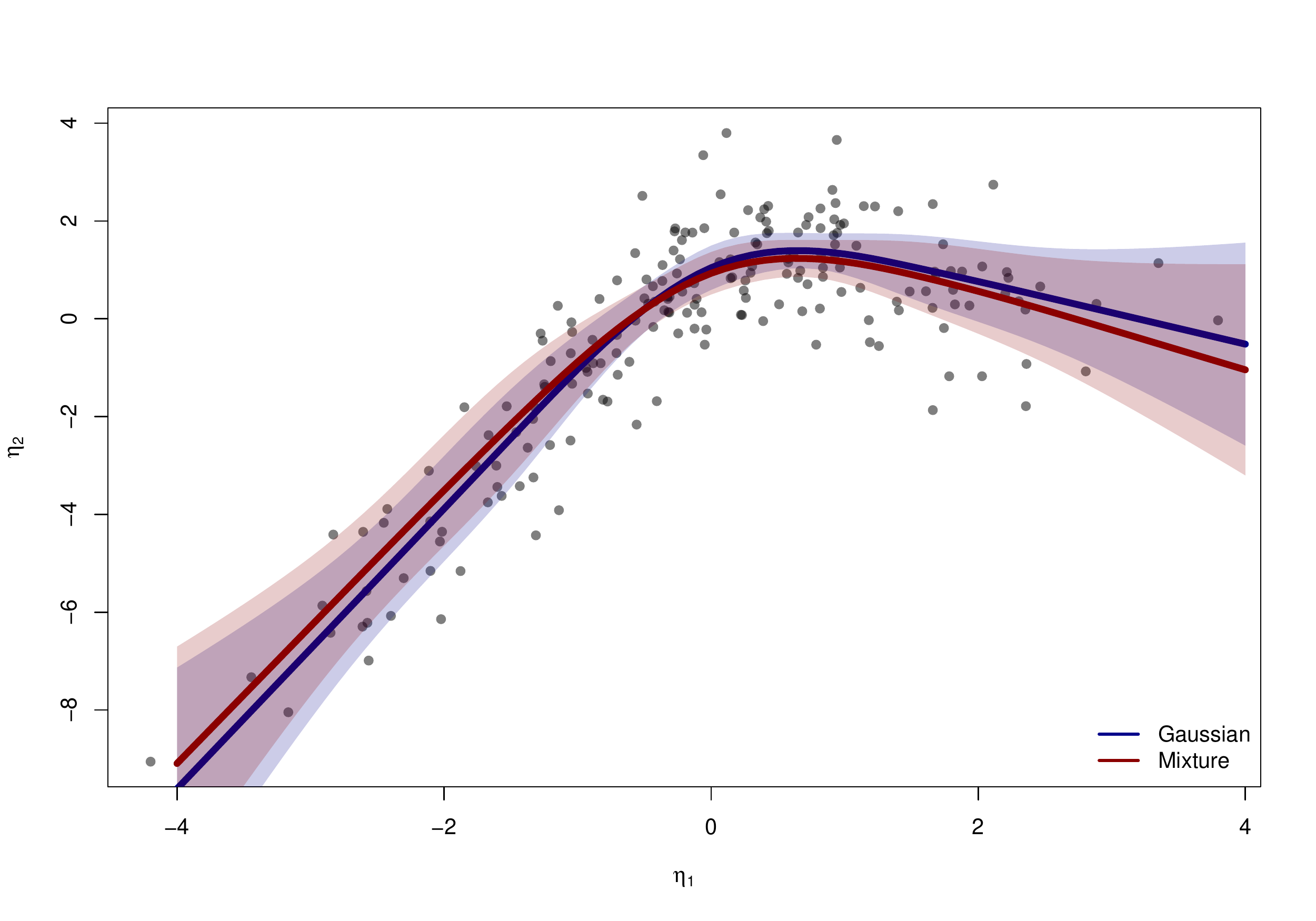}
\caption{Model fit with point-wise 95\% confidence limits where the
  measurement model of \(\eta_{1}\) is assumed to follow a Gaussian
  distribution or two-component Gaussian mixture distribution. label:fig:mixturefit}
\end{figure*}

\clearpage
\section{Simulation study}
\label{appendix:simulation}

In the following we study the structural equation model defined by the
stage 1 model:
\begin{align}
\xi &= \gamma_1Z + \widetilde{\zeta}, \qquad Z \sim \mathcal{N}(0,1),  \quad  
X_{ij} = \xi + \widetilde{\epsilon}_{j}, \quad j=1,2,3  
\end{align}
and the stage 2 model given by
\begin{align*}
\eta = \beta^T\varphi(\xi) + \gamma_2Z + \zeta, \qquad
Y_{j} = \eta + \epsilon_{j}, \quad j=1,2,3
\end{align*}
with mutually independent residual terms
\(\widetilde{\zeta}, \zeta, \widetilde{\epsilon}_{j}, \epsilon_{j}, \
j=1,2,3\). The distribution of these terms was varied throughout the
simulations.  The parameters of primary interest are the structural
parameters, \(\beta\), defining the association between the two latent
variables \(\xi\) and \(\eta\).

The source code for all the simulations are available in the
\texttt{twostage} branch of the \texttt{lava} git repository (commit
hash: \texttt{53e2e18}): 
\url{https://github.com/kkholst/lava/tree/twostage/inst/simulations}.

\subsection{Correctly specified model}

We first conducted a simulation study to explore
the properties of our estimator in correctly
specified models. Data was generated from a quadratic structural model
\begin{align*}
 \eta = \beta_{0} + \beta_{1}\xi + \beta_{2}\xi^{2} + \gamma_2Z + \zeta,
\end{align*}
with all residuals \(\widetilde{\zeta}_i, \zeta_i, \widetilde{\epsilon}_{ij}, \epsilon_{ij},
 j=1,2,3\) being standard normal and
with \(\beta_{0}=1, \beta_{1}=1, \beta_{2}=0.5\). We compared the \twostage estimator
(with and with-out mixture model extension) to 2SLS estimation,
methods of moments estimation, and approximate ML based on a Laplace
approximation as well as Gaussian Adaptive Quadrature with 9
quadrature points. For the mixture \twostage estimator we considered
only two-component mixtures of the form
\(\widetilde{\zeta} \sim \pi\mathcal{N}(\mu_1,\sigma^2) +
(1-\pi)\mathcal{N}(\mu_2,\sigma^2)\).

\subsubsection{Gaussian model without covariates (\(n=500\))}
\label{sec:sim:gaussian500}
Here we studied the situation when no covariates were included in neither
the data generating model (\(\gamma_1=\gamma_2=0\)) and the in
the estimated models. Sample-size was set to \(n=500\) and all
residual terms where normal distributed. See Table~\ref{tab:sim1}.

\begin{table}[!h]
  \centering
\begin{tabular}[t]{llllllll}
  \toprule
  \bfseries{} & \bfseries{} & \bfseries{Mean} & \bfseries{SD} & 
  \bfseries{SE} & \(\tfrac{\text{\bfseries{SE}}}{\text{\bfseries{SD}}}\) & \bfseries{Cov.} & \bfseries{RMSE}\\
  \midrule
 & \twostage & 1.013 & 0.483 & 0.495 & 1.027 & 0.950 & 0.483\\

 & \twostage mixture & 1.075 & 0.511 & 0.544 & 1.065 & 0.951 & 0.517\\

 & 2SLS & 1.031 & 0.978 & 0.893 & 0.913 & 0.940 & 0.978\\

 & 2SLS robust & 1.031 & 0.978 & 0.910 & 0.931 & 0.939 & 0.978\\

 & 2SMM & 1.108 & 0.623 & {} & {} & {} & 0.633\\

 & 2SMM robust & 1.141 & 0.702 & {} & {} & {} & 0.716\\

 & Laplace & 1.161 & 0.428 & 0.420 & 0.982 & 0.948 & 0.457\\

\multirow{-8}{*}{\raggedright\arraybackslash \(\bm{\beta_1=1}\)} & AGQ9 & 1.030 & 0.382 & 0.390 & 1.023 & 0.956 & 0.383\\

  \midrule
 & \twostage & 0.501 & 0.094 & 0.097 & 1.026 & 0.950 & 0.094\\

 & \twostage mixture & 0.515 & 0.100 & 0.107 & 1.075 & 0.952 & 0.101\\

 & 2SLS & 0.507 & 0.173 & 0.147 & 0.851 & 0.920 & 0.173\\

 & 2SLS robust & 0.507 & 0.173 & 0.162 & 0.933 & 0.934 & 0.173\\

 & 2SMM & 0.520 & 0.116 & {} & {} & {} & 0.117\\

 & 2SMM robust & 0.526 & 0.131 & {} & {} & {} & 0.133\\

 & Laplace & 0.532 & 0.085 & 0.083 & 0.973 & 0.959 & 0.091\\

\multirow{-8}{*}{\raggedright\arraybackslash \(\bm{\beta_2=0.5}\)} & AGQ9 & 0.506 & 0.075 & 0.076 & 1.013 & 0.951 & 0.075\\
  
  \bottomrule
\end{tabular}
\caption{\label{tab:sim1} Simulations from quadratic model without any
  covariates and  Gaussian distribution of all residual
  terms. Sample-size was \(n=500\) and the number of replications of
  the simulation study was 1,000.
}
\end{table}

\clearpage
\subsubsection{Gaussian model without covariates (\(n=1,000\))}
\label{sec:sim:gaussian1000}
Same setup as in the previous section but with sample size increased
to \(n=1,000\). See Table~\ref{tab:sim2}.

\begin{table}[!h]
  \centering
\begin{tabular}[t]{llllllll}
  \toprule
  \bfseries{} & \bfseries{} & \bfseries{Mean} & \bfseries{SD} & 
  \bfseries{SE} & \(\tfrac{\text{\bfseries{SE}}}{\text{\bfseries{SD}}}\) & \bfseries{Cov.} & \bfseries{RMSE}\\
  \midrule
 & \twostage & 0.996 & 0.371 & 0.349 & 0.942 & 0.940 & 0.371\\

 & \twostage mixture & 1.031 & 0.393 & 0.424 & 1.079 & 0.941 & 0.394\\

 & 2SLS & 0.975 & 0.647 & 0.606 & 0.938 & 0.938 & 0.647\\

 & 2SLS robust & 0.975 & 0.647 & 0.631 & 0.975 & 0.948 & 0.647\\

 & 2SMM & 1.031 & 0.431 & {} & {} & {} & 0.432\\

 & 2SMM robust & 1.042 & 0.466 & {} & {} & {} & 0.468\\

 & Laplace & 1.127 & 0.306 & 0.292 & 0.957 & 0.939 & 0.331\\

\multirow{-8}{*}{\raggedright\arraybackslash \(\bm{\beta_1=1}\)} & AGQ9 & 1.002 & 0.277 & 0.272 & 0.984 & 0.947 & 0.277\\

  \midrule
 & \twostage & 0.499 & 0.072 & 0.068 & 0.944 & 0.933 & 0.072\\

 & \twostage mixture & 0.507 & 0.077 & 0.084 & 1.095 & 0.934 & 0.077\\

 & 2SLS & 0.496 & 0.115 & 0.100 & 0.864 & 0.919 & 0.115\\

 & 2SLS robust & 0.496 & 0.115 & 0.112 & 0.973 & 0.942 & 0.115\\

 & 2SMM & 0.506 & 0.081 & {} & {} & {} & 0.081\\

 & 2SMM robust & 0.508 & 0.088 & {} & {} & {} & 0.088\\

 & Laplace & 0.526 & 0.060 & 0.057 & 0.958 & 0.939 & 0.065\\

\multirow{-8}{*}{\raggedright\arraybackslash \(\bm{\beta_2=0.5}\)} & AGQ9 & 0.501 & 0.054 & 0.053 & 0.986 & 0.943 & 0.054\\

  \bottomrule
\end{tabular}
\caption{\label{tab:sim2}
  Simulations from quadratic model without any
  covariates and  Gaussian distribution of all residual
  terms. Sample-size was \(n=1,000\) and the number of replications of
  the simulation study was 1,000.}
\end{table}

\clearpage
\subsubsection{Gaussian model with a covariate (\(n=500\))}
\label{sec:sim:gaussianX500}
Here we included a covariate \(Z\) in the model with
\(\gamma_1=\gamma_2=1\). The sample-size was set to \(n=500\) and all
residual terms where normal distributed. See Table~\ref{tab:sim3}.
As the methods of moments estimator is not immediately available for
estimating models with covariates \citep{wall2000estimation} we omitted it from this simulation.

\begin{table}[!h]
  \centering
\begin{tabular}[t]{llllllll}
  \toprule
  \bfseries{} & \bfseries{} & \bfseries{Mean} & \bfseries{SD} & 
  \bfseries{SE} & \(\tfrac{\text{\bfseries{SE}}}{\text{\bfseries{SD}}}\) & \bfseries{Cov.} & \bfseries{RMSE}\\
  \midrule
 & \twostage & 0.996 & 0.249 & 0.254 & 1.017 & 0.954 & 0.249\\

 & \twostage mixture & 0.991 & 0.248 & 0.256 & 1.029 & 0.956 & 0.249\\

 & 2SLS & 1.016 & 0.412 & 0.398 & 0.966 & 0.944 & 0.413\\

 & 2SLS robust & 1.016 & 0.412 & 0.407 & 0.987 & 0.946 & 0.413\\



 & Laplace & 1.041 & 0.227 & 0.227 & 1.002 & 0.949 & 0.231\\

\multirow{-6}{*}{\raggedright\arraybackslash \(\bm{\beta_1=1}\)} & AGQ9 & 1.000 & 0.219 & 0.222 & 1.010 & 0.945 & 0.219\\

  \midrule
 & \twostage & 0.499 & 0.048 & 0.050 & 1.034 & 0.954 & 0.048\\

 & \twostage mixture & 0.501 & 0.049 & 0.051 & 1.043 & 0.952 & 0.049\\

 & 2SLS & 0.503 & 0.077 & 0.061 & 0.798 & 0.894 & 0.077\\

 & 2SLS robust & 0.503 & 0.077 & 0.076 & 0.987 & 0.949 & 0.077\\



 & Laplace & 0.510 & 0.045 & 0.045 & 0.986 & 0.946 & 0.046\\

\multirow{-6}{*}{\raggedright\arraybackslash \(\bm{\beta_2=0.5}\)} & AGQ9 & 0.500 & 0.043 & 0.043 & 0.998 & 0.942 & 0.043\\

  \bottomrule
\end{tabular}
\caption{\label{tab:sim3}
  Simulations from quadratic with one
  covariate (\(\gamma_1=1, \gamma_2=1\)) and  Gaussian distribution of all residual
  terms. Sample-size was \(n=500\) and the number of replications of
  the simulation study was 1,000.}
\end{table}

\clearpage
\subsubsection{Gaussian model with a covariate (\(n=1,000\))}
\label{sec:sim:gaussianX1000}
Same setup as in the previous section but with sample size increased
to \(n=1,000\). See Table~\ref{tab:sim4}.

\begin{table}[!h]
  \centering
\begin{tabular}[t]{llllllll}
  \toprule
  \bfseries{} & \bfseries{} & \bfseries{Mean} & \bfseries{SD} & 
  \bfseries{SE} & \(\tfrac{\text{\bfseries{SE}}}{\text{\bfseries{SD}}}\) & \bfseries{Cov.} & \bfseries{RMSE}\\
  \midrule
 & \twostage & 0.997 & 0.178 & 0.181 & 1.017 & 0.950 & 0.178\\

 & \twostage mixture & 0.993 & 0.178 & 0.182 & 1.027 & 0.950 & 0.178\\

 & 2SLS & 1.000 & 0.286 & 0.278 & 0.973 & 0.940 & 0.286\\

 & 2SLS robust & 1.000 & 0.286 & 0.291 & 1.019 & 0.951 & 0.286\\



 & Laplace & 1.039 & 0.157 & 0.162 & 1.026 & 0.958 & 0.162\\

\multirow{-6}{*}{\raggedright\arraybackslash \(\bm{\beta_1=1}\)} & AGQ9 & 0.998 & 0.151 & 0.156 & 1.033 & 0.961 & 0.151\\

  \midrule
 & \twostage & 0.500 & 0.035 & 0.036 & 1.026 & 0.953 & 0.035\\

 & \twostage mixture & 0.501 & 0.035 & 0.036 & 1.040 & 0.950 & 0.035\\

 & 2SLS & 0.500 & 0.054 & 0.042 & 0.789 & 0.871 & 0.054\\

 & 2SLS robust & 0.500 & 0.054 & 0.054 & 1.009 & 0.950 & 0.054\\



 & Laplace & 0.510 & 0.031 & 0.032 & 1.041 & 0.953 & 0.032\\

\multirow{-6}{*}{\raggedright\arraybackslash \(\bm{\beta_2=0.5}\)} & AGQ9 & 0.500 & 0.029 & 0.030 & 1.050 & 0.958 & 0.029\\

  \bottomrule
\end{tabular}
\caption{\label{tab:sim4}
  Simulations from quadratic with one
  covariate (\(\gamma_1=1, \gamma_2=1\)) and  Gaussian distribution of all residual
  terms. Sample-size was \(n=1,000\) and the number of replications of
  the simulation study was 1,000.}
\end{table}

\clearpage
\subsubsection{Exponential model}

We also assessed the performance of the model with an exponential effect
\(\eta = \beta_{1}\xi + \beta_2\exp(\xi) + \zeta.\)
The consistency of the
2SLS estimator relies on the assumption that the function \(\varphi\) can be
decomposed additively as
\(\varphi(\xi) = \varphi(x_1-\widetilde{\epsilon}_1) = g_1(x_1) + g_2(x_1,\widetilde{\epsilon}_1)\) and with access to
instrumental variables (indicators \(X_i\)) that should be
uncorrelated with the second term, \(g_2\). This is not possible
with the exponential function, however in a neighbourhood around zero 
\(\exp(x_1-\widetilde{\epsilon}_1) \approx \exp(x_1) + \exp(x_1)(x_1-\widetilde{\epsilon}_1)\), which suggests a
first order approximate 2SLS estimator with  \(g_1(x_1) = \exp(x_1)\).

The results of the simulation with an exponential transformation are
summarized in Table \ref{tab:simexp} (\(n=500\)). As expected we see that the
\twostage estimator is unbiased while this is no longer the case for
the 2SLS estimator. Also, the coverage of the \twostage estimator
remains close to the nominal level, however with more extreme choices
of the parameters (simulations not shown) we did observe that the
sample size needed to be increased to obtain correct coverage levels.

\begin{table}[!h]
  \centering
\begin{tabular}[t]{llllllll}
  \toprule
  \bfseries{} & \bfseries{} & \bfseries{Mean} & \bfseries{SD} & 
  \bfseries{SE} & \(\tfrac{\text{\bfseries{SE}}}{\text{\bfseries{SD}}}\) & \bfseries{Cov.} & \bfseries{RMSE}\\
  \midrule
 & \twostage & -0.005 & 0.197 & 0.196 & 0.994 & 0.950 & 0.197\\

  \multirow{-2}{*}{\raggedright\arraybackslash \(\bm{\beta_1=0}\)}
              & 2SLS & 0.074 & 0.739 & 0.670 & 0.906 & 0.906 & 0.742\\
  \midrule
 & \twostage & 0.303 & 0.173 & 0.172 & 0.997 & 0.905 & 0.173\\

  \multirow{-2}{*}{\raggedright\arraybackslash \(\bm{\beta_1=0.3}\)}
              & 2SLS & 0.151 & 0.398 & 0.342 & 0.859 & 0.698 & 0.425\\

  \bottomrule
\end{tabular}
\caption{\label{tab:simexp} Performance of the Gaussian two-stage estimator (\twostage),
  under an exponential model,
  \(\E(\eta\mid\xi) = \beta_0 + \beta_1\xi + \beta_2\exp(\xi)\),
  with true  parameters \(\beta_1=0\) and \(\beta_2=0.3\)) where all assumptions hold.
  The \twostage estimator is compared to a first order approximate 2SLS estimator.}
\end{table}

\subsection{Robustness}

Here we explored the properties of the estimators in misspecified
models.  First we examined data generating mechanisms, where the
conditional distribution of the latent variable \(\xi\) was not Gaussian
but followed a mixture distribution, i.e.,
\begin{align*}
  \widetilde{\zeta} \sim \pi\mathcal{N}(\mu_1,\sigma^2) + (1-\pi)\mathcal{N}(\mu_2,\sigma^2).
\end{align*}

The above simulation setup corresponds exactly to the assumptions of
our mixture model extension, so to test the robustness of the
extension we also included a study where \(\widetilde{\zeta}\) followed a
uniform distribution with mean zero and variance one, and a simulation
where the residuals of the indicators, \(\widetilde{\epsilon}_i\) 
followed a uniform distribution. 

\clearpage
\subsubsection{Mixture model without covariates (\(n=500\))}
Same as Section \ref{sec:sim:gaussian500} with \(\widetilde{\zeta}\)
following a Gaussian mixture distribution with
\(\pi=\tfrac{1}{4}, \sigma^2=1, \mu_1=0, \mu_2=3\).  See
Table~\ref{tab:sim5}.

\begin{table}[!h]
  \centering
\begin{tabular}[t]{llllllll}
  \toprule
  \bfseries{} & \bfseries{} & \bfseries{Mean} & \bfseries{SD} & 
  \bfseries{SE} & \(\tfrac{\text{\bfseries{SE}}}{\text{\bfseries{SD}}}\) & \bfseries{Cov.} & \bfseries{RMSE}\\
  \midrule
 & \twostage & 1.358 & 0.126 & 0.123 & 0.980 & 0.162 & 0.379\\

 & \twostage mixture & 1.002 & 0.148 & 0.148 & 0.999 & 0.946 & 0.148\\

 & 2SLS & 1.014 & 0.229 & 0.280 & 1.220 & 0.981 & 0.230\\

 & 2SLS robust & 1.014 & 0.229 & 0.231 & 1.008 & 0.945 & 0.230\\

 & 2SMM & 0.998 & 0.163 & {} & {} & {} & 0.163\\

  \multirow{-6}{*}{\raggedright\arraybackslash \(\bm{\beta_1=1}\)}
              & 2SMM robust & 0.996 & 0.166 & {} & {} & {} & 0.166\\

  \midrule
 & \twostage & 0.378 & 0.041 & 0.041 & 1.001 & 0.164 & 0.129\\

 & \twostage mixture & 0.499 & 0.053 & 0.054 & 1.023 & 0.948 & 0.053\\

 & 2SLS & 0.495 & 0.076 & 0.073 & 0.957 & 0.921 & 0.077\\

 & 2SLS robust & 0.495 & 0.076 & 0.078 & 1.020 & 0.937 & 0.077\\

 & 2SMM & 0.500 & 0.056 & {} & {} & {} & 0.056\\

  \multirow{-6}{*}{\raggedright\arraybackslash \(\bm{\beta_2=0.5}\)}
              & 2SMM robust & 0.501 & 0.058 & {} & {} & {} & 0.058\\

  \bottomrule
\end{tabular}
\caption{\label{tab:sim5} Simulations from quadratic model without any
  covariates and the latent variable of the stage one model following
  a Gaussian mixture distribution.  Sample-size was \(n=500\) and
  the number of replications of the simulation study was 1,000.}
\end{table}

\clearpage
\subsubsection{Mixture model without covariates (\(n=1,000\))}
Same as Section \ref{sec:sim:gaussian1000} with \(\widetilde{\zeta}\)
following a Gaussian mixture distribution with
\(\pi=\tfrac{1}{4}, \sigma^2=1, \mu_1=0, \mu_2=3\).  See
Table~\ref{tab:sim6}.

\begin{table}[!h]
  \centering
\begin{tabular}[t]{llllllll}
  \toprule
  \bfseries{} & \bfseries{} & \bfseries{Mean} & \bfseries{SD} & 
  \bfseries{SE} & \(\tfrac{\text{\bfseries{SE}}}{\text{\bfseries{SD}}}\) & \bfseries{Cov.} & \bfseries{RMSE}\\
  \midrule
 & \twostage & 1.349 & 0.088 & 0.086 & 0.976 & 0.021 & 0.360\\

 & \twostage mixture & 1.000 & 0.104 & 0.103 & 0.988 & 0.948 & 0.104\\

 & 2SLS & 1.012 & 0.169 & 0.195 & 1.149 & 0.978 & 0.170\\

 & 2SLS robust & 1.012 & 0.169 & 0.162 & 0.956 & 0.936 & 0.170\\

 & 2SMM & 1.000 & 0.112 & {} & {} & {} & 0.112\\

  \multirow{-6}{*}{\raggedright\arraybackslash \(\bm{\beta_1=1}\)}
              & 2SMM robust & 0.999 & 0.114 & {} & {} & {} & 0.114\\

  \midrule
 & \twostage & 0.380 & 0.030 & 0.029 & 0.976 & 0.025 & 0.123\\

 & \twostage mixture & 0.499 & 0.038 & 0.038 & 0.990 & 0.947 & 0.038\\

 & 2SLS & 0.496 & 0.056 & 0.051 & 0.903 & 0.915 & 0.056\\

 & 2SLS robust & 0.496 & 0.056 & 0.055 & 0.976 & 0.929 & 0.056\\

 & 2SMM & 0.500 & 0.040 & {} & {} & {} & 0.040\\

  \multirow{-6}{*}{\raggedright\arraybackslash \(\bm{\beta_2=0.5}\)}
              & 2SMM robust & 0.500 & 0.041 & {} & {} & {} & 0.041\\

  \bottomrule
\end{tabular}
\caption{\label{tab:sim6}
  Simulations from quadratic model without any
  covariates and the latent variable of the stage one model following
  a Gaussian mixture distribution.  Sample-size was \(n=1,000\) and
  the number of replications of the simulation study was 1,000.
}
\end{table}

\clearpage
\subsubsection{Mixture model with a covariate (\(n=500\))}
Same as Section \ref{sec:sim:gaussianX500} with \(\widetilde{\zeta}\)
following a Gaussian mixture distribution with
\(\pi=\tfrac{1}{4}, \sigma^2=1, \mu_1=0, \mu_2=3\).  See
Table~\ref{tab:sim7}.

\begin{table}[!h]
  \centering
\begin{tabular}[t]{llllllll}
  \toprule
  \bfseries{} & \bfseries{} & \bfseries{Mean} & \bfseries{SD} & 
  \bfseries{SE} & \(\tfrac{\text{\bfseries{SE}}}{\text{\bfseries{SD}}}\) & \bfseries{Cov.} & \bfseries{RMSE}\\
  \midrule
 & \twostage & 1.102 & 0.100 & 0.100 & 1.001 & 0.817 & 0.142\\

 & \twostage mixture & 1.003 & 0.100 & 0.103 & 1.027 & 0.956 & 0.100\\

 & 2SLS & 1.006 & 0.167 & 0.199 & 1.193 & 0.977 & 0.167\\


  \multirow{-4}{*}{\raggedright\arraybackslash \(\bm{\beta_1=1}\)}
              & 2SLS robust & 1.006 & 0.167 & 0.160 & 0.962 & 0.940 & 0.167\\

  \midrule
 & \twostage & 0.446 & 0.034 & 0.033 & 0.974 & 0.599 & 0.064\\

 & \twostage mixture & 0.498 & 0.037 & 0.037 & 0.992 & 0.943 & 0.037\\

 & 2SLS & 0.497 & 0.055 & 0.046 & 0.833 & 0.889 & 0.055\\

  \multirow{-4}{*}{\raggedright\arraybackslash \(\bm{\beta_2=0.5}\)}
              & 2SLS robust & 0.497 & 0.055 & 0.053 & 0.961 & 0.929 & 0.055\\
  \bottomrule
\end{tabular}
\caption{\label{tab:sim7}
  Simulations from quadratic with one
  covariate (\(\gamma_1=1, \gamma_2=1\)) and the
  latent variable of the stage one model following
  a Gaussian mixture distribution
  Sample-size was \(n=500\) and the number of replications of
  the simulation study was 1,000.}
\end{table}

\clearpage
\subsubsection{Mixture model with a covariate (\(n=1,000\))}
Same as Section \ref{sec:sim:gaussianX1000} with \(\widetilde{\zeta}\)
following a Gaussian mixture distribution with
\(\pi=\tfrac{1}{4}, \sigma^2=1, \mu_1=0, \mu_2=3\).  See
Table~\ref{tab:sim8}.

\begin{table}[!h]
  \centering
\begin{tabular}[t]{llllllll}
  \toprule
  \bfseries{} & \bfseries{} & \bfseries{Mean} & \bfseries{SD} & 
  \bfseries{SE} & \(\tfrac{\text{\bfseries{SE}}}{\text{\bfseries{SD}}}\) & \bfseries{Cov.} & \bfseries{RMSE}\\
  \midrule
 & \twostage & 1.095 & 0.072 & 0.070 & 0.979 & 0.736 & 0.119\\

 & \twostage mixture & 0.997 & 0.073 & 0.072 & 0.994 & 0.940 & 0.073\\

 & 2SLS & 1.002 & 0.117 & 0.140 & 1.198 & 0.982 & 0.117\\


  \multirow{-4}{*}{\raggedright\arraybackslash \(\bm{\beta_1=1}\)}
 & 2SLS robust & 1.002 & 0.117 & 0.113 & 0.968 & 0.933 & 0.117\\

  \midrule
 & \twostage & 0.448 & 0.023 & 0.023 & 1.023 & 0.398 & 0.057\\

 & \twostage mixture & 0.500 & 0.025 & 0.026 & 1.034 & 0.953 & 0.025\\

 & 2SLS & 0.498 & 0.038 & 0.032 & 0.846 & 0.896 & 0.038\\


  \multirow{-4}{*}{\raggedright\arraybackslash \(\bm{\beta_2=0.5}\)}
              & 2SLS robust & 0.498 & 0.038 & 0.037 & 0.988 & 0.932 &
                                                                    0.038\\
  
  \bottomrule
\end{tabular}
\caption{\label{tab:sim8}
    Simulations from quadratic with one
  covariate (\(\gamma_1=1, \gamma_2=1\)) and the
  latent variable of the stage one model following
  a Gaussian mixture distribution
  Sample-size was \(n=1,000\) and the number of replications of
  the simulation study was 1,000.
}
\end{table}

\clearpage
\subsubsection{\(\widetilde{\zeta} \sim
  \mathcal{U}(-\frac{\sqrt12}{2}, \frac{\sqrt12}{2}) \  (n=500)\)}
Same as Section \ref{sec:sim:gaussian500} with \(\widetilde{\zeta}\)
following a zero-mean uniform distribution with variance 1.  See
Table~\ref{tab:sim9}.

\begin{table}[!h]
  \centering
\begin{tabular}[t]{llllllll}
  \toprule
  \bfseries{} & \bfseries{} & \bfseries{Mean} & \bfseries{SD} & 
  \bfseries{SE} & \(\tfrac{\text{\bfseries{SE}}}{\text{\bfseries{SD}}}\) & \bfseries{Cov.} & \bfseries{RMSE}\\
  \midrule
 & \twostage & 0.999 & 0.098 & 0.098 & 0.995 & 0.949 & 0.098\\

 & \twostage mixture (2) & 1.000 & 0.109 & 0.109 & 1.005 & 0.944 & 0.109\\

 & \twostage mixture (3) & 0.993 & 0.107 & 0.110 & 1.031 & 0.942 & 0.107\\

 & 2SLS & 1.006 & 0.123 & 0.121 & 0.986 & 0.950 & 0.123\\

 & 2SLS robust & 1.006 & 0.123 & 0.128 & 1.040 & 0.957 & 0.123\\

 & 2SMM & 1.005 & 0.111 & {} & {} & {} & 0.111\\

  \multirow{-7}{*}{\raggedright\arraybackslash \(\bm{\beta_1=1}\)}
              & 2SMM robust & 1.005 & 0.110 & {} & {} & {} & 0.110\\

  \midrule
 & \twostage & 0.310 & 0.079 & 0.079 & 1.005 & 0.327 & 0.205\\

 & \twostage mixture (2) & 0.493 & 0.151 & 0.154 & 1.016 & 0.932 & 0.151\\

 & \twostage mixture (3) & 0.474 & 0.145 & 0.168 & 1.162 & 0.940 & 0.147\\

 & 2SLS & 0.488 & 0.260 & 0.228 & 0.879 & 0.928 & 0.260\\

 & 2SLS robust & 0.488 & 0.260 & 0.242 & 0.932 & 0.931 & 0.260\\

 & 2SMM & 0.555 & 0.185 & {} & {} & {} & 0.193\\

  \multirow{-7}{*}{\raggedright\arraybackslash \(\bm{\beta_1=0.5}\)}
              & 2SMM robust & 0.541 & 0.203 & {} & {} & {} & 0.207\\

  \bottomrule
\end{tabular}
\caption{\label{tab:sim9}  
  Simulations from quadratic model without any
  covariates and the latent variable of the stage one model following
  a uniform distribution.  Sample-size was \(n=500\) and
  the number of replications of the simulation study was 1,000.}
\end{table}

\clearpage
\subsubsection{\(\widetilde{\zeta} \sim
  \mathcal{U}(-\frac{\sqrt12}{2}, \frac{\sqrt12}{2}) \  (n=1,000)\)}
Same as Section \ref{sec:sim:gaussian1000} with \(\widetilde{\zeta}\)
following a zero-mean uniform distribution with variance 1.  See
Table~\ref{tab:sim10}.

\begin{table}[!h]
  \centering
\begin{tabular}[t]{llllllll}
  \toprule
  \bfseries{} & \bfseries{} & \bfseries{Mean} & \bfseries{SD} & 
  \bfseries{SE} & \(\tfrac{\text{\bfseries{SE}}}{\text{\bfseries{SD}}}\) & \bfseries{Cov.} & \bfseries{RMSE}\\
  \midrule
 & \twostage & 0.998 & 0.069 & 0.069 & 0.998 & 0.948 & 0.069\\

 & \twostage mixture (2) & 0.998 & 0.075 & 0.075 & 1.007 & 0.957 & 0.075\\

 & \twostage mixture (3) & 0.994 & 0.075 & 0.076 & 1.012 & 0.953 & 0.075\\

 & 2SLS & 1.003 & 0.085 & 0.082 & 0.958 & 0.944 & 0.086\\

 & 2SLS robust & 1.003 & 0.085 & 0.087 & 1.015 & 0.959 & 0.086\\

 & 2SMM & 1.001 & 0.078 & {} & {} & {} & 0.078\\

  \multirow{-7}{*}{\raggedright\arraybackslash \(\bm{\beta_1=1}\)}
              & 2SMM robust & 1.001 & 0.077 & {} & {} & {} & 0.077\\

  \midrule
 & \twostage & 0.310 & 0.054 & 0.056 & 1.030 & 0.102 & 0.198\\

 & \twostage mixture (2) & 0.485 & 0.101 & 0.104 & 1.027 & 0.944 & 0.102\\

 & \twostage mixture (3) & 0.485 & 0.100 & 0.107 & 1.071 & 0.960 & 0.101\\

 & 2SLS & 0.489 & 0.160 & 0.149 & 0.930 & 0.923 & 0.161\\

 & 2SLS robust & 0.489 & 0.160 & 0.159 & 0.994 & 0.934 & 0.161\\

 & 2SMM & 0.544 & 0.124 & {} & {} & {} & 0.131\\

  \multirow{-7}{*}{\raggedright\arraybackslash \(\bm{\beta_1=0.5}\)}
              & 2SMM robust & 0.520 & 0.126 & {} & {} & {} & 0.128\\

  \bottomrule
\end{tabular}
\caption{\label{tab:sim10}  
  Simulations from quadratic model without any
  covariates and the latent variable of the stage one model following
  a uniform distribution.  Sample-size was \(n=1,000\) and
  the number of replications of the simulation study was 1,000.
}
\end{table}

\clearpage

\subsubsection{\(\widetilde \epsilon_j \sim
  \mathcal{U}(-\frac{\sqrt12}{2}, \frac{\sqrt12}{2}) \  (n=500)\)}
Same as Section \ref{sec:sim:gaussian1000} with \(\widetilde{\epsilon}_j, j=1,2,3\)
following a zero-mean uniform distribution with variance 1.  See
Table~\ref{tab:sim11}.

\begin{table}[!h]
  \centering
\begin{tabular}[t]{llllllll}
  \toprule
  \bfseries{} & \bfseries{} & \bfseries{Mean} & \bfseries{SD} & 
  \bfseries{SE} & \(\tfrac{\text{\bfseries{SE}}}{\text{\bfseries{SD}}}\) & \bfseries{Cov.} & \bfseries{RMSE}\\
  \midrule
 & \twostage & 0.998 & 0.107 & 0.107 & 0.997 & 0.949 & 0.107\\

 & \twostage mixture (2) & 1.002 & 0.117 & 0.118 & 1.012 & 0.942 & 0.117\\

 & \twostage mixture (3) & 0.997 & 0.115 & 0.115 & 0.997 & 0.945 & 0.115\\

 & 2SLS & 1.003 & 0.138 & 0.116 & 0.844 & 0.915 & 0.138\\

 & 2SLS robust & 1.003 & 0.138 & 0.131 & 0.951 & 0.953 & 0.138\\

 & 2SMM & 1.004 & 0.117 & {} & {} & {} & 0.117\\

  \multirow{-7}{*}{\raggedright\arraybackslash \(\bm{\beta_1=1}\)}
              & 2SMM robust & 1.005 & 0.118 & {} & {} & {} & 0.118\\

  \midrule
 & \twostage & 0.506 & 0.089 & 0.089 & 0.997 & 0.948 & 0.089\\

 & \twostage mixture (2) & 0.527 & 0.106 & 0.108 & 1.018 & 0.961 & 0.109\\

 & \twostage mixture (3) & 0.506 & 0.105 & 0.106 & 1.017 & 0.946 & 0.105\\

 & 2SLS & 0.498 & 0.131 & 0.102 & 0.781 & 0.873 & 0.131\\

 & 2SLS robust & 0.498 & 0.131 & 0.124 & 0.944 & 0.909 & 0.131\\

 & 2SMM & 0.525 & 0.107 & {} & {} & {} & 0.110\\

  \multirow{-7}{*}{\raggedright\arraybackslash \(\bm{\beta_1=0.5}\)}
              & 2SMM robust & 0.519 & 0.115 & {} & {} & {} & 0.117\\

  \bottomrule
\end{tabular}
\caption{\label{tab:sim11}
    Simulations from quadratic model without any
  covariates and the residuals of the manifest variables of the stage one model following
  a uniform distribution.  Sample-size was \(n=500\) and
  the number of replications of the simulation study was 1,000.  
}
\end{table}

\clearpage
\subsubsection{\(\widetilde \epsilon_j \sim
  \mathcal{U}(-\frac{\sqrt12}{2}, \frac{\sqrt12}{2}) \  (n=1,000)\)}
Same as Section \ref{sec:sim:gaussian1000} with \(\widetilde{\epsilon}_j, j=1,2,3\)
following a zero-mean uniform distribution with variance 1.  See
Table~\ref{tab:sim12}.

\begin{table}[!h]
  \centering
\begin{tabular}[t]{llllllll}
  \toprule
  \bfseries{} & \bfseries{} & \bfseries{Mean} & \bfseries{SD} & 
  \bfseries{SE} & \(\tfrac{\text{\bfseries{SE}}}{\text{\bfseries{SD}}}\) & \bfseries{Cov.} & \bfseries{RMSE}\\
  \midrule
 & \twostage & 0.997 & 0.075 & 0.075 & 1.002 & 0.945 & 0.075\\

 & \twostage mixture (2) & 0.998 & 0.082 & 0.082 & 1.000 & 0.935 & 0.082\\

 & \twostage mixture (3) & 0.996 & 0.081 & 0.082 & 1.016 & 0.939 & 0.081\\

 & 2SLS & 0.997 & 0.090 & 0.081 & 0.894 & 0.923 & 0.090\\

 & 2SLS robust & 0.997 & 0.090 & 0.091 & 1.014 & 0.951 & 0.090\\

 & 2SMM & 1.000 & 0.079 & {} & {} & {} & 0.079\\

  \multirow{-7}{*}{\raggedright\arraybackslash \(\bm{\beta_1=1}\)}
              & 2SMM robust & 1.000 & 0.079 & {} & {} & {} & 0.079\\

  \midrule
 & \twostage & 0.506 & 0.063 & 0.063 & 0.996 & 0.951 & 0.063\\

 & \twostage mixture (2) & 0.521 & 0.072 & 0.076 & 1.055 & 0.956 & 0.075\\

 & \twostage mixture (3) & 0.511 & 0.072 & 0.076 & 1.065 & 0.957 & 0.072\\

 & 2SLS & 0.499 & 0.088 & 0.070 & 0.801 & 0.886 & 0.088\\

 & 2SLS robust & 0.499 & 0.088 & 0.088 & 0.996 & 0.937 & 0.088\\

 & 2SMM & 0.521 & 0.070 & {} & {} & {} & 0.073\\

  \multirow{-7}{*}{\raggedright\arraybackslash \(\bm{\beta_1=0.5}\)}
              & 2SMM robust & 0.512 & 0.074 & {} & {} & {} & 0.074\\

  \bottomrule
\end{tabular}
\caption{\label{tab:sim12}
  Simulations from quadratic model without any
  covariates and the residuals of the manifest variables of the stage one model following
  a uniform distribution.  Sample-size was \(n=1,000\) and
  the number of replications of the simulation study was 1,000.
}
\end{table}

\clearpage

\clearpage
\subsection{Larger measurement error}
\label{sec:simlargemeasurementerror}

Same as Section \ref{sec:sim:gaussianX500} with \(\widetilde{\epsilon}_j, j=1,2,3\)
following a zero-mean normal distribution with variance 2.  See
Table~\ref{tab:simlargemeasure}. We observe that also in this case
with increased measurement error in the stage one model, the \twostage
estimator performs well. Here we also include results on the parameter
estimates of the intercepts and covariate effect in the stage two
model. Bias in the 2SLS intercept is observed.

\scriptsize
\begin{table}[!h]
  \centering
\resizebox{30em}{!}{%
  \begin{tabular}[t]{llllllll}
  \toprule
    &  & \bfseries{Mean} & \bfseries{SD} & 
  \bfseries{SE} & \(\tfrac{\text{\bfseries{SE}}}{\text{\bfseries{SD}}}\) & \bfseries{Cov.} & \bfseries{RMSE}\\
  \midrule
 & 2SSEM & 1.008 & 0.479 & 0.474 & 0.989 & 0.958 & 0.479\\

 & 2SSEM mixture & 0.926 & 0.492 & 0.519 & 1.054 & 0.958 & 0.498\\

 & 2SLS & 0.024 & 0.784 & 0.966 & 1.232 & 0.860 & 1.252\\

 & 2SLS robust & 0.024 & 0.784 & 0.780 & 0.995 & 0.719 & 1.252\\

 & Laplace & 1.253 & 0.392 & 0.390 & 0.996 & 0.922 & 0.466\\

\multirow{-6}{*}{\raggedright\arraybackslash \(\bm{\beta_0=1}\)} & AGQ9 & 0.999 & 0.372 & 0.375 & 1.008 & 0.952 & 0.372\\
  \midrule
 & 2SSEM & 0.992 & 0.340 & 0.345 & 1.014 & 0.951 & 0.340\\

 & 2SSEM mixture & 0.975 & 0.340 & 0.353 & 1.038 & 0.954 & 0.341\\

 & 2SLS & 1.017 & 0.711 & 0.695 & 0.978 & 0.951 & 0.711\\

 & 2SLS robust & 1.017 & 0.711 & 0.690 & 0.971 & 0.946 & 0.711\\

 & Laplace & 1.059 & 0.291 & 0.289 & 0.994 & 0.956 & 0.297\\

\multirow{-6}{*}{\raggedright\arraybackslash \(\bm{\beta_Z=1}\)} & AGQ9 & 0.994 & 0.276 & 0.280 & 1.015 & 0.944 & 0.276\\
\midrule
 & 2SSEM & 0.992 & 0.340 & 0.345 & 1.014 & 0.951 & 0.340\\

 & 2SSEM mixture & 0.975 & 0.340 & 0.353 & 1.038 & 0.954 & 0.341\\

 & 2SLS & 1.017 & 0.711 & 0.695 & 0.978 & 0.951 & 0.711\\

 & 2SLS robust & 1.017 & 0.711 & 0.690 & 0.971 & 0.946 & 0.711\\

 & Laplace & 1.059 & 0.291 & 0.289 & 0.994 & 0.956 & 0.297\\

\multirow{-6}{*}{\raggedright\arraybackslash \(\bm{\beta_1=1}\)} & AGQ9 & 0.994 & 0.276 & 0.280 & 1.015 & 0.944 & 0.276\\
\midrule
 & 2SSEM & 0.498 & 0.068 & 0.071 & 1.029 & 0.948 & 0.069\\

 & 2SSEM mixture & 0.501 & 0.069 & 0.073 & 1.060 & 0.954 & 0.069\\

 & 2SLS & 0.503 & 0.132 & 0.109 & 0.822 & 0.893 & 0.132\\

 & 2SLS robust & 0.503 & 0.132 & 0.128 & 0.970 & 0.937 & 0.132\\

 & Laplace & 0.511 & 0.060 & 0.059 & 0.977 & 0.946 & 0.061\\

\multirow{-6}{*}{\raggedright\arraybackslash \(\bm{\beta_2=0.5}\)} & AGQ9 & 0.499 & 0.057 & 0.057 & 1.000 & 0.936 & 0.057\\
\bottomrule
  \end{tabular}
}
\caption{\label{tab:simlargemeasure} Simulations from quadratic model
  with one covariate (\(\gamma_1=1, \gamma_2=1\)) and Gaussian
  distribution of all residual terms. Sample-size was \(n=500\) and
  the number of replications of the simulation study was 1,000. All
  residual terms have variance 1 except
  \(\widetilde{\epsilon}_ij\sim\mathcal{N}(0,2)\).}
\end{table}
\normalsize

\clearpage
\subsection{Non-parametric estimation}
\label{sec:simsupnonpar}

  To study the estimator in a non-parametric setting we also simulated
  data from the latent variable model with measurement models as defined
  in the previous sections but with the unknown functional relationship
  between the two measurement models given by
  \begin{align*}
    \phi(\xi; \bm{\beta}) = \beta_1\xi + \beta_2\xi^2 + \sin(\beta_3\xi)
  \end{align*}
  
  As a benchmark we compared results with the estimator proposed by
  \citep{kelava2017} and the corresponding Matlab
  implementation\footnotemark\footnotetext{\url{https://github.com/tifasch/nonparametric/tree/ead709097d6}}.
  Here the number of equidistant spline knots were chosen by
  dividing the simulated data into a single test and training dataset
  of equal size and choosing the spline basis (degrees of freedom
  varying from 1 to 11) as the one that minimized the RMSE evaluated
  in the test dataset.  We noted that slightly better results were obtained for our
  two-stage estimator when the hyper-parameters (spline
  knots) were chosen using 5-fold cross validation. To
  make the results more comparable we, however, adopted the same method  
  for choosing the degrees of freedom for the spline using the exact
  same split of the testing and training data.

\begin{figure}[h!]
  \centering
  \includegraphics[width=0.4\textwidth]{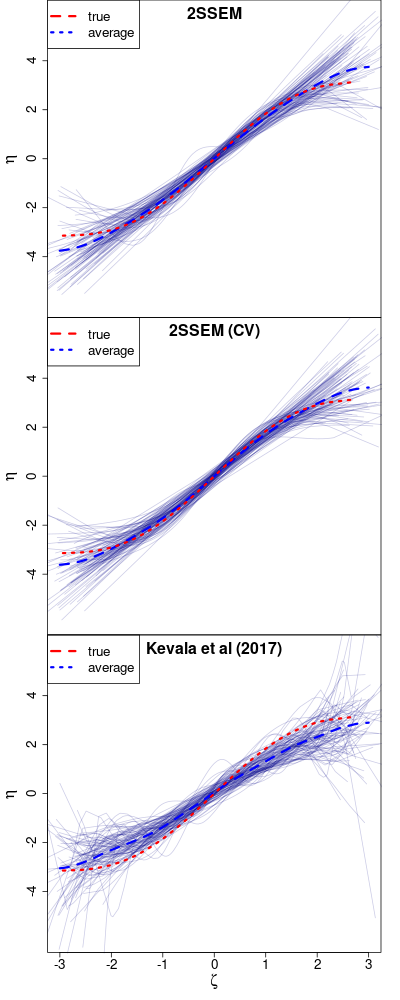}
  \caption{\label{fig:nonpar}Comparison of the two-stage estimator and
  the estimation procedure of
  \cite{kelava2017}. The dashed red line is the
  true association between \(\xi\) and \(\eta\) and the dotted blue
  line is the average estimated association over the 100
  replications. The transparent lines show the estimated spline for
  each of the replications.}

\end{figure}

  In each simulation, \(r=1,\ldots,100\), we simulated \(n=200\)
  observations, and for each estimator we calculated   
  \begin{align*}
      \text{RMSE}_{r} = \left\{\sum_{i=1}^n \left[\phi(\xi_{i};
    \bm{\beta}) - \tilde\phi(\xi_i; \widehat{\bm{\gamma}}_r)\right]^2\right\}^{1/2},
  \end{align*}
  where \(\bm{\beta}\) denotes the true parameter and
  \(\widehat{\bm{\gamma}}_{r}\) is the estimated parameters of the
  spline model, i.e.,
  \(\widehat{\eta} = \tilde\phi(\xi; \widehat{\bm{\gamma}}_r) =
  \bm{B}(\xi)\widehat{\bm{\gamma}}_r\), where \(\bm{B}(\xi)\) is the
  spline basis design matrix.  See Table~\ref{tab:nonpar} and
  Figure~\ref{fig:nonpar} where the estimates are shown from the
  scenario where \(\beta_1=1, \beta_2=0, \beta_3=1\) and with
  \(\widetilde{\zeta},\zeta\in\mathcal{N}(0,1)\).

\begin{table}
  \centering
\begin{tabular}[t]{rrrllclll}
  \toprule
  & & & \multicolumn{2}{c}{\bf{Model}} & \phantom{A} & \multicolumn{3}{c}{\bf{RMSE}} \\
   \cmidrule(lr){4-5}  \cmidrule(lr){7-9}
\textbf{\(\bm{\beta_1}\)} & \textbf{\(\bm{\beta_2}\)} & \textbf{\(\bm{\beta_3}\)}
      & \(\widetilde{\bm{\zeta}}\) & \(\bm{\zeta}\) &  & \textbf{Kevala} & \textbf{\twostage} & \textbf{\twostage-CV(5)}\\
\midrule
1 & 0 & 1 & \(\mathcal{N}(0,1)\) & \(\mathcal{N}(0,1)\) & & 0.314 & 0.112 & 0.098\\
1 & 0 & 1 & \(\mathcal{U}(-6,6)\) & \(\mathcal{U}(-6,6)\) & & 0.933 & 0.608 & 0.585\\
1 & 0 & 1 & \(\mathcal{GM}(-4,4,0.5)\) &\(\mathcal{N}(0,1)\) & & 0.200 & 0.143 & 0.101\\
\addlinespace
0 & 1 & 3 & \(\mathcal{N}(0,1)\) & \(\mathcal{N}(0,1)\) & & 1.177 & 0.827 & 0.625\\
0 & 1 & 3 & \(\mathcal{U}(-6,6)\) & \(\mathcal{U}(-6,6)\) & & 3.614 & 1.988 & 1.860\\
0 & 1 & 3 & \(\mathcal{GM}(-4,4,0.5)\) &\(\mathcal{N}(0,1)\) & & 6.322 & 1.658 & 1.590\\
\bottomrule
\end{tabular}
\caption{\label{tab:nonpar} Comparison of the 2SMM estimator and
  the estimation procedure of \cite{kelava2017}.}
\end{table}

\end{document}